\documentclass[twocolumn,aps,prd,showpacs,letterpaper,superscriptaddress]{revtex4}
\usepackage{graphicx}
\graphicspath{{figures/for_archive}}

\begin{document}

\hyphenation{Ka-pi-tul-nik}

\title{Constraints on Yukawa-Type Deviations from Newtonian Gravity at 20 Microns}
\author{S.~J.~Smullin}
\altaffiliation{Present address: Physics Department, Princeton University, Princeton, NJ 08544, USA}
\author{A.~A.~Geraci}
\author{D.~M.~Weld}
\affiliation{Department of Physics, Stanford University, Stanford, CA 94305, USA}
\author{J.~Chiaverini}
\altaffiliation{Present address: Los Alamos National Laboratory, MS D454, Los Alamos, NM 87545, USA}
\affiliation{National Institute of Standards and Technology, Boulder, Colorado 80305, USA}
\author{S.~Holmes}
\affiliation{Department of Statistics, Stanford University, Stanford, CA 94305, USA}
\author{A.~Kapitulnik}
\email{aharonk@stanford.edu}
\affiliation{Department of Physics, Stanford University, Stanford, CA 94305, USA}
\affiliation{Department of Applied Physics, Stanford University, Stanford, CA 94305, USA}

\date{\today}
\pacs{04.80.Cc}
\begin{abstract}
Recent theories of physics beyond the standard model have predicted deviations from Newtonian gravity at short distances. In order to test these theories, we have a built an apparatus that can measure attonewton-scale forces between gold masses separated by distances on the order of 25~$\mu$m. A micromachined silicon cantilever was used as the force sensor, and its displacement was measured with a fiber interferometer. We have used our measurements to set bounds on the magnitude $\alpha$ and length scale $\lambda$ of Yukawa-type deviations from Newtonian gravity; our results presented here yield the best experimental limit in the range of $\lambda=6$--20~$\mu$m. \end{abstract}
\maketitle

\section{Introduction}
\subsection{Motivation}
The theories of Newton and Einstein are powerful and accurate in describing the observed natural world. In spite of this, gravity still presents several theoretical challenges, including the gauge hierarchy problem, the cosmological constant problem, and the lack of a quantum description of gravity. Many theories of physics beyond the standard model, in particular those theories that attempt to unify the standard model with gravity, predict the existence of extra dimensions, exotic particles, or new forces that could cause a mass coupling in addition to the Newtonian gravitational potential at short distances. Of particular interest to us have been some recent theories that predict new forces in a range measurable by tabletop experiments \cite{add1,add2,sundrum,savasmoduli,savasradion,ignatios,dilaton,beane}. In many cases, the modification to Newtonian gravity is predicted to be a Yukawa-type potential, arising either from coupling to massive particles or the compactification of extra dimensions.  With this addition, the gravitational potential between two masses $m_1$ and $m_2$ separated by distance $r$ is predicted to be of the form:
\begin{equation}\label{yukeqn}V(r)=-G{{m_1m_2}\over{r}}\left [1+\alpha e^{-(r/\lambda)}\right].\end{equation}

\noindent Here, $G$ is Newton's constant, $\alpha$ is the strength of the new potential as compared to the Newtonian gravitational potential, and $\lambda$ is its range. 

\begin{figure}
\includegraphics[width=\columnwidth]{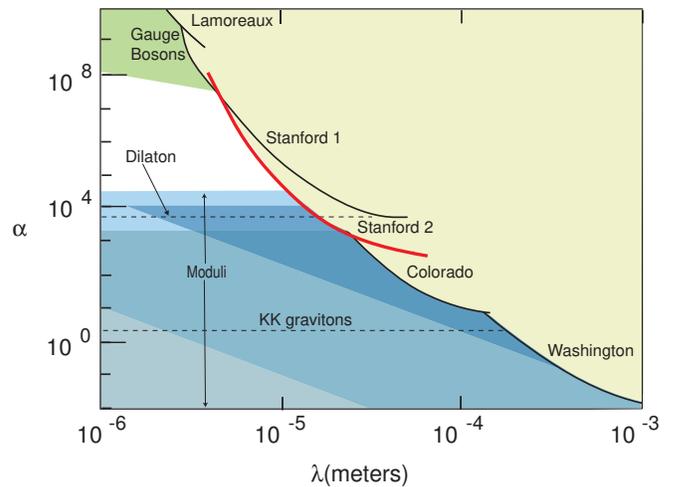}
\caption{[color online] Experimental results (solid lines) and theoretical predictions shown in $\alpha$-$\lambda$ space. The area to the upper right of the experimental lines (from Refs.~\cite{lamoreaux,lamoreauxalphlam,frogs,pricenature,adelberger04} and our latest results from this paper) shows where Yukawa-type deviations from Newtonian gravity have been excluded. The line labeled ``Stanford 1'' gives results from \cite{frogs}; the line labeled ``Stanford 2'' shows the results described in this paper. Dashed lines show roughly the predictions for the dilaton \cite{dilaton} and the first Kaluza-Klein mode of two simply-compactified extra dimensions as described in Ref.~\cite{add2}. Shaded regions to the left of the experimental lines show predictions for moduli and gauge bosons from Ref.~\cite{andysavas}. \label{paramspace}}
\end{figure}

With these ideas in mind, we have constructed a device to measure attonewton-scale forces between masses separated by distances on the order of 25~$\mu$m \cite{jacthesis,frogs,SJSmoriond,SJSslac,SJSthesis}. Our experiment was designed to measure Yukawa-type forces in the range of $\lambda=$ 5--50~$\mu$m with as small a value of $|\alpha|$ as possible. This range of parameters is relevant to recent theoretical predictions \cite{add2,savasmoduli,dilaton,beane} and is complementary to recent experimental attempts to test these new ideas \cite{frogs,lamoreaux,adelberger04,pricenature}. Our results are reported in terms of an upper bound on $\alpha(\lambda)$ of a Yukawa potential that could be consistent with our data; we present results at the $95\%$ confidence level. Unless otherwise stated, all measurements and techniques described apply to the Cooldown 04 experiment, from which the new $\alpha(\lambda)$ bound was derived. These latest results yield close to an order of magnitude improvement over our previous bounds on the Yukawa potential at distances on the order of 20~$\mu$m \cite{frogs}. Fig.~\ref{paramspace} summarizes our results together with some of the theoretical predictions and other recent experimental results.

\subsection{Overview}
To measure very short-range forces, we used masses of size comparable to the distance between them. A gold prism was attached to the end of a single-crystal silicon cantilever. A drive mass, comprising a gold meander pattern embedded in a silicon substrate to create an alternating pattern of gold and silicon bars, was mounted on a piezoelectric bimorph at a distance from the cantilever-mounted test mass. The face-to-face vertical separation between the masses was $\sim$~25~$\mu$m, limited in part by the presence of a stiff, metallized silicon nitride shield membrane between the masses. 

\begin{figure}
\includegraphics[width=0.8\columnwidth]{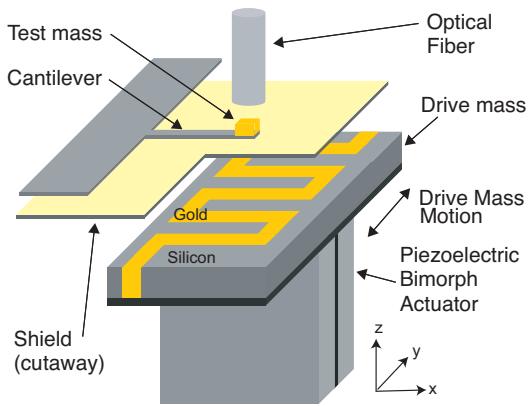}
\caption{[color online] Schematic (not to scale) showing the drive mass below the cantilever that bears the gold test mass. A gold-coated silicon nitride shield membrane separates the masses.\label{schematic}}
\end{figure}

As diagrammed in Fig.~\ref{schematic}, the drive mass was oscillated along the $y$-direction underneath the test mass, maintaining the vertical ($z$-) separation between the masses. Due to the differing mass densities of the gold and silicon bars, an alternating gravitational field was created at the test mass location. The drive mass was oscillated at a subharmonic of the resonant frequency of the cantilever. Due to the geometry of the drive mass and the amplitude of oscillation, any gravitational coupling between the masses would thus create a force on the cantilever at harmonics of this drive frequency, including the cantilever's resonant frequency. The motion of the cantilever on resonance was measured using a fiber interferometer; from this measurement, the force between the masses was deduced. 

Thermal noise provided a limit on the measurement of cantilever motion. To minimize this limit, cantilevers were fabricated to have small spring constants and high quality factors in vacuum. The thermal noise limit in this experiment was approximately $2.5\times 10^{-16}$~N/$\sqrt{\rm{Hz}}$, at cryogenic temperatures in vacuum. 

In order to be able to test and characterize the apparatus by measuring a known force much larger than any expected gravitational force, a magnetic analog to the gravitational experiment was built into the apparatus. Magnetic test masses were fabricated with a nickel layer on them. An electrical current passed through the drive mass meander created a spatially-varying magnetic field that would couple to the magnetic moment of the test mass and drive the cantilever. 

For both magnetic and gravitational tests, the force was measured as a function of the equilibrium $y$-position between the drive mass and the test mass. Any coupling between masses would show a distinct periodicity in the measured force as a function of this $y$-position. By comparing our measurements to predictions from finite element analysis (FEA), a bound on Yukawa-type deviations was derived.

The separation of the signal frequency from the drive frequency, the shield between the masses, and the use of non-magnetic test masses for the gravitational measurement reduced or eliminated many possible sources of non-gravitational background. The geometry of the apparatus provided an important degree of freedom that permitted a discrimination of true coupling between the masses from certain electrical or mechanical backgrounds. 

\subsection{Organization of the Paper}
This paper primarily describes the apparatus and results from the experiment labeled Cooldown 04. Cooldown 01 is described in Refs.~\cite{frogs,jacthesis}, Cooldown 02 is described in Sec.~\ref{testssec} of this paper, and Cooldown 03 is described elsewhere\cite{SJSslac,SJSthesis}. 

In the second section of this paper, the apparatus is described in detail. The following sections describe our finite element analysis, the experimental method, and the averaging used to convert the raw data to a force measurement. The sixth section of this paper describes the magnetic experiment used to test the apparatus and the measurement of thermal noise. In the final sections, our experimental results, fitting techniques, error analysis, and the resulting bound on $\alpha(\lambda)$ are presented, followed by conclusions and a discussion of future prospects for this experiment.

\section{Apparatus}
\subsection{Cantilever} The cantilevers used in this experiment were fabricated from single-crystal silicon using standard micromachining techniques \cite{jacthesis}. The cantilevers were fabricated from $\left<100\right>$ Si, oriented in the $\left<110\right>$ direction; they were 50~$\mu$m wide, 250~$\mu$m long, and 0.33~$\mu$m thick, yielding an expected spring constant $k=0.005$~N/m \cite{sarid,E_Si_size}. 

The resonant frequency of a mass-loaded cantilever is determined by the spring constant $k$ and the mass of the test mass $m_t$:\begin{equation}\omega_0^2=k/m_t ,\end{equation} where $\omega_0$ is the angular frequency of the first bending mode of the cantilever. The spring constant of each cantilever was deduced from the resonant frequency, which was measured very precisely, and the mass of the test mass, discussed in the next section. Adding the test mass reduced the resonant frequency of the cantilever to $\sim 300$~Hz. As found from the resonant frequency, the addition of the test mass increased the spring constant to $\sim$~0.007~N/m; this increase was due to the shortened effective length of the cantilever once the test mass was attached.
 
Cantilevers exhibit a Lorentzian transfer function between driving force and amplitude; when driven at the resonant frequency $f_0$ by force $F$, the maximum displacement $x$ at the center of mass of the test mass is \begin{equation} \label{acforceeqn}x(f_0)=F(f_0)Q/k ,\end{equation} where $Q$ is the quality factor. The quality factors of cantilevers used in this experiment were found to be as high as 80000 in vacuum and at cryogenic temperatures. 

The energy from the thermal environment provides a constant series of random-phase impulses at all frequencies to the device. The resulting motion of the cantilever shows the Lorentzian transfer function, with a force spectral density $S_f$ on resonance of \begin{equation}S_f=\sqrt{4 k k_B T/Q\omega_0}\label{thnoiseeqn} ,\end{equation} where $T$ is the cantilever temperature and $k_B$ is the Boltzmann constant. We used floppy (low spring constant), high quality factor cantilevers at low temperatures to reduce this limit. Force measurements were averaged as long as needed to see a signal above noise or as long as was practical. 

\subsection{Test Mass}\label{tmsec}
To fabricate test masses, gold was deposited (using a thermal evaporator) into molds of plasma-etched silicon. After polishing of the top surface of evaporated gold, the silicon was dissolved to release the test masses. To make magnetic test masses, 1000~{\AA} of Ni was evaporated before the gold. The test masses were designed to be prisms $(50\times 50 \times 30)$~$\mu$m$^3$ in size; with these dimensions, a test mass could be affixed to a cantilever with one of its larger faces flush against the cantilever and aligned with the end of the cantilever.

        \begin{table}
        \begin{center}
        \begin{minipage}{\columnwidth}
        \begin{center}
        \caption{Dimensions of the test mass. }
        \smallskip
        \begin{tabular} { l  r  r  r}
        \hline
        \hline
          Parameter & Value & Error &Units\\
        \hline		
         Width				& 51			&1.5		&$\mu$m\\
         Length 				& 51			&1.5		&$\mu$m\\
         Rectangular height 		&32			&1.5		&$\mu$m\\
         Rounded volume 		&7800		&3000	&$\mu$m$^3$\\
         Missing volume from side face		&4000		&1500	&$\mu$m$^3$	\\
         	Missing volume due to porosity &2.5	&2.5		&$\%$\\
	Density of gold			&19.3		&0		&g/cm$^3$\\
         \hline	
        \textbf{\textit{Total {\bf Mass of test mass}}}	&1.64	&0.13	&$\mu$g\\
        \hline
        \hline
        \end{tabular}
        \label{masserror}
        \end{center}
        \end{minipage}
        \end{center}
        \end{table}

A group of thirty test masses was weighed (in sets of five or six) with a microbalance to determine the average mass. Because many of these test masses were imperfect specimens, this measurement was only used as a guide. After weighing, test masses were examined under a scanning electron microscope (SEM) and etched by a focused ion beam (FIB) to determine more precisely the typical dimensions and porosity of the test masses. All the nonmagnetic test masses were fabricated together; those examined in the FIB were assumed to be representative of the entire batch, including the one used in Cooldown~04. The SEM showed the test masses to be slightly larger than intended (as confirmed by the microbalance measurements), with a rounded face where the evaporated gold was polished. Etching with the FIB showed the evaporated gold to have some porosity near the side faces of the test mass. 

\begin{figure}
\includegraphics[width=0.5\columnwidth]{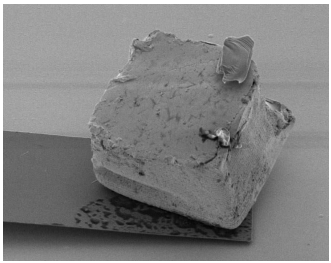}
\caption{Scanning electron micrograph of the test mass and cantilever used in Cooldown 04. A piece of dust is visible on one corner of the test mass. The cantilever is 50~$\mu$m wide.\label{testmasspic}}
\end{figure}

After data acquisition, the test mass and cantilever used in Cooldown 04 were examined under an SEM (Fig.~\ref{testmasspic}). It was found that the test mass was mounted with its rounded side closer to the cantilever and with the top flat side tilted at 0.35 rad with respect to the cantilever in the $y$-$z$~plane. Estimates of the size and uncertainty in the dimensions of the test mass were derived from examination of this test mass and of the larger group of masses from the same fabrication batch. SEM images of the Cooldown 04 test mass showed that the side face that was further away from the cantilever (due to the tilt of the test mass) was partially recessed; an estimate of this missing volume was included in the calculation of the mass.  Error on these estimates was due to error in the SEM measurements and uncertainty in the exact shape of the rounded part on the polished side. The density of the gold was assumed to be the bulk density of 19.3 g/cm$^3$; the observed porosity was included in volume and mass uncertainty. The dimensions and density of the test mass and the experimental uncertainties in these parameters are shown in Tbl.~\ref{masserror}.  The mass given in Tbl.~\ref{masserror} agrees well (within the given experimental uncertainties) with the mass deduced from a comparison of the resonant frequencies of the cantilever with the mass and neighboring cantilevers without masses.

\subsection{Drive Mass}
The drive mass was also fabricated by evaporating gold into a mold of silicon. After polishing, the silicon substrate was diced into dies approximately 1.8~mm~$\times$~1.3~mm in size. 

Gold and silicon have differing electrical conductivities, as well as differing mass densities. To eliminate the possibility of a periodic coupling between the masses due to the Casimir force or charging of the drive mass, the pattern of the drive mass was buried beneath a thin ground plane. To bury the drive mass pattern, the polished side of the die was mounted on a quartz backplane, which became the bottom of the drive mass. The bulk of the remaining silicon on the top was removed, leaving a layer less than 2~$\mu$m thick. On top of this layer of silicon was deposited a thin layer of aluminum oxide (for electrical insulation) and 1000~{\AA} of gold on top of an adhesion layer of titanium. This gold film was continued around to the side of the drive mass, where electrical contact was made to a ground wire glued on with silver epoxy. This thin ground plane masked variations in the Casimir force without notably affecting the varying gravitational field of the drive mass.  

\begin{figure}
\includegraphics[width=0.75\columnwidth]{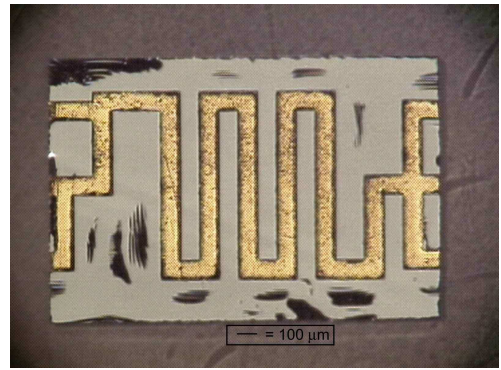}
\caption{[color online] Optical micrograph of the polished side of a drive mass showing the meander pattern. In the experiment, the polished side of the drive mass was facing away from the test mass.\label{dmphoto}}
\end{figure}

As shown in Fig.~\ref{dmphoto}, the main part of the drive mass pattern comprised five sets of gold and silicon bars, each 1~mm long and 100~$\mu$m in each of the cross-sectional dimensions. The rest of the drive mass pattern provided leads to which electrical contact could be made for grounding the meander (for the gravitational experiment) or for passing electrical current through the meander (for the magnetic experiment).

A drive mass similar to the one used for measurement was etched by the FIB and examined under the SEM. On the polished side of the die, each gold bar had a band of indeterminate composition on either side, where the polishing created a wedge of silicon, gold, and polishing grit mixed together. This wedge extended 10--20~$\mu$m into the gold bar at each gold/silicon boundary, tapering off at a depth of $\sim 10$~$\mu$m into the bar; this band of indeterminate composition thus comprised a small part of the ($100\times 100$)~$\mu$m$^2$ cross-section of each gold bar. After final fabrication and mounting of the buried drive mass, this polished side of the drive mass was facing away from the test mass. Because the Yukawa potentials in question are short-range, this imperfection at a distance of $>100$~$\mu$m away from the test mass would have little effect on the results. In the analysis, this imperfection was taken as a small error on the bulk density of the gold in the drive mass. It was assumed that the drive mass had porosity on the sides of the gold bars similar to the amount observed in the etched test mass.  

\subsection{Shield Membrane}
The test mass and cantilever were isolated from electrostatic and Casimir excitations by a shield membrane between the cantilever wafer die and the drive mass. The cantilever was held within a silicon wafer die approximately 1~cm$^2$ in size. This die was glued to another silicon die, which was etched into a frame  bearing a 3-$\mu$m thick membrane of silicon nitride across an area of 5.2~mm~$\times$~2.8~mm. The entire shield wafer die, including the membrane, was coated with gold on both sides, with a ground wire attached to one corner of the die. Due to the geometry and the tensile stress in the membrane \cite{snf_nitridestress,sin_stress_a,sin_stress_b}, the membrane was expected to be much stiffer than the cantilever. 

If there was some force between the drive mass and the shield at $f_0$, the shield would move at $f_0$. The shield motion could drive the cantilever capacitively or via the Casimir force. However, because of the stiffness of the shield and the separation of the drive frequency from the signal frequency, any interaction between the shield and the drive mass would likely be too small to make the shield deflect enough to drive the cantilever a measurable amount on resonance. 

\subsection{Piezoelectric Bimorph Actuator}
A piezoelectric bimorph actuator (hereafter referred to as the ``bimorph'') was used to move the drive mass longitudinally underneath the test mass. The bimorph was driven at a subharmonic of the cantilever resonant frequency $f_0$; the particular subharmonic (either $f_0/3$ or $f_0/4$) was chosen so the drive frequency was below the resonance of the bimorph but high enough to gain a resonant enhancement in the amplitude and a reduction of nonlinearities in the bimorph motion. At low temperatures, self-heating of the bimorph, resonant enhancement of the motion, and driving voltages larger than the room temperature limits for the device allowed an amplitude of 100--125~$\mu$m of motion at a drive frequency $f_d$ of 90--120~Hz. Finite element analysis showed that the magnitude of the time-varying force at $3f_d$ from the Newtonian and any Yukawa potential between masses would vary as a function of the bimorph amplitude, with the maximum occurring at a bimorph amplitude of $\sim 135$~$\mu$m. 

The bimorph was secured at its base in a brass and Cirlex \cite{cirlex} clamp. On one side of the bimorph was glued a capacitive electrode facing a counter-electrode mounted on the clamp. On the other side of the bimorph was mounted a small mirror. Prior to being mounted in the probe, the bimorph was  calibrated. Room-temperature calibration of the bimorph was performed by measuring the rms capacitance between the two electrodes with a given driving voltage on the bimorph. The corresponding amount of motion was determined by reflecting a laser beam off the mirror onto a linear CCD array. At low temperatures, the amplitude of the bimorph motion was determined from a measurement of the rms capacitance. 

Uncertainties in the bimorph motion were tabulated through the linear fits used to compare the capacitance measurement {\it in situ} to the calibration. The total experimental uncertainty in the amplitude of motion was 10~$\mu$m, mostly a result of uncertainty in the original measurement of the laser spot position from the CCD array. 

A small ground cap was glued on top of the bimorph, to help isolate the shield membrane from the large ac voltages used to drive the bimorph. The drive mass was glued on top of this ground plane. 

For purposes of the calibration, the bimorph bending shape was considered to be an arc with constant radius along the length of the bimorph. For our analysis, it was assumed that the drive mass was moving purely in the horizontal plane; in fact as the bimorph bends, the drive mass will tip up at the ends, changing the vertical separation between the test mass and the drive mass below it. Depending on the alignment of the masses, the resulting change in the vertical separation between masses may be as much as 2~$\mu$m over the course of one period of drive mass motion. This effect was not expected to substantially change the signal and a full modeling of it has been left for future work. 

\subsection{Vibration Isolation}
Measurement at the thermal noise limit required mechanical excitation at the cantilever tip to be less than 1~{\AA} (rms) on resonance; the mechanical excitation at the base of the cantilever was required to be reduced a factor of $Q$ beyond this.  In order to make a sensitive measurement, it was crucial to ensure that the bimorph was not simply mechanically exciting the cantilever. Even though the bimorph was moved at a subharmonic of the resonant frequency of the cantilever, nonlinearities in the bimorph motion (as in any piezoelectric device) meant that some component (typically a few percent) of its motion was at $f_0$; this necessitated vibration isolation between the bimorph stage and the cantilever wafer stage. 

The bimorph and cantilever wafer were separated by two simple spring-mass vibration isolation stages. These stages each had resonant frequencies of $\sim$~2.5~Hz, calculated to yield roughly six orders of magnitude of attenuation at the bimorph frequency and eight orders of magnitude of attenuation at the signal frequency $f_0$. 

The electrical wires and the optical fiber that ran the length of the probe could have, if not loose enough, shorted out the vibration isolation system. Measurements of the cantilever at low temperatures proved to be the best test of the vibration isolation system. The mechanical coupling between bimorph and cantilever could be assessed by measuring the motion of the cantilever with the bimorph moving at a large ($\sim 1$~mm) vertical separation between masses, with the signal frequency on and off resonance of the cantilever.  Such couplings were found to vary over the course of one cooldown, with no clear indication of failure of the vibration isolation.

External vibration isolation protected the fragile parts of the probe from human mechanical disturbances when the drive mass was only microns away from the shield membrane. During data-acquisition, the entire cryogenic system was suspended from a thick concrete ceiling by springs, yielding a resonant frequency for the entire system of close to 2~Hz.

\subsection{Capacitive Position and Tilt Sensors}
Capacitive position and tilt sensors were used to indicate the relative position between the bimorph and wafer stages, so that the masses could be aligned to maximize the gravitational force between them. The capacitive position sensor (CPS) was based on the design described in Ref.~\cite{cps}. A quadrature pattern of gold electrodes was patterned on a quartz substrate and these electrodes were mounted on the bimorph stage. Above this, on the wafer stage, was mounted a single sense electrode. An ac voltage was applied to the quadrature electrodes and the current was read from the counter-electrode via a dual-phase lockin amplifier. With a phase shift of 90 degrees between the signals applied to each of the four quadrature electrodes, the two channels of the lockin amplifier provided readings linear in the relative $x$ and $y$ positions between the bimorph stage and the wafer stage. With the signal applied in-phase to each of the quadrature electrodes, the signal provided an indication of $1/z_c$, where $z_c$ is the vertical separation between the sense and quadrature electrodes. 

\begin{figure}
\includegraphics[width=\columnwidth]{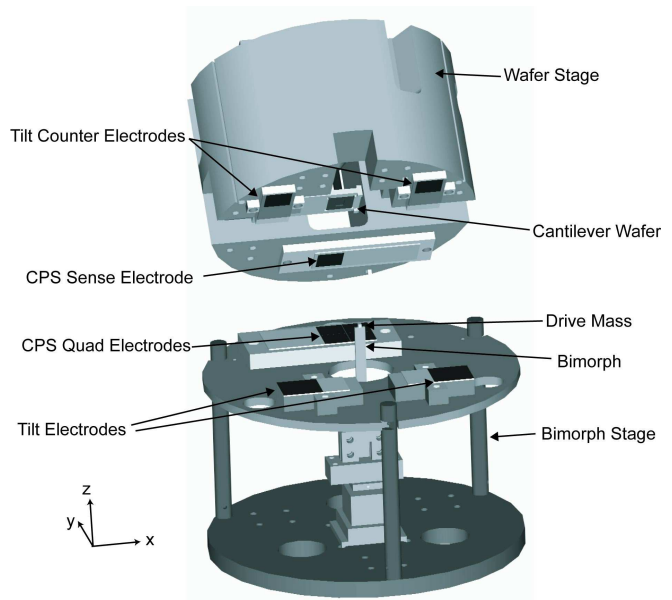}
\caption{Drawing of the bottom half of the probe. The bimorph stage and wafer stage are shown with the wafer stage raised and rotated about the $x$-axis to reveal the geometry of the capacitive sensors. In practice, the two electrodes of a given sensor were separated by less than 1 mm in the $z$-direction. The three sensors were separated by $\sim$ 4 cm in the $x$-$y$ plane. The pads used for making electrical contact to the capacitive sensors are omitted from this drawing.\label{cpsgeom}}
\end{figure}

The tilt sensors were simpler, each comprising an electrode mounted on the bimorph stage and a counter-electrode on the wafer stage above. In each case, an ac voltage was applied to one electrode, with the current read from the corresponding counter-electrode via a lockin amplifier. The tilt sensors provided an indication of the vertical separation between the stages at two other locations in addition to the CPS. The geometry of the three capacitive sensors is shown in Fig.~\ref{cpsgeom}.

Together, the capacitive sensors provided an indication of the relative position and tilt between the stages, encompassing five degrees of freedom. The relative rotation between the stages about the $z$-axis was fixed during probe assembly.

The capacitive sensors were calibrated at room temperature by raising the bimorph stage with a three-axis positioner and recording the micrometer readings from the $z$-axis of the positioner along with the capacitive readings. The CPS was also calibrated in $x$ and $y$ using a similar technique. With the probe under vacuum, the three-axis positioning stage tilted due to atmospheric pressure; this tilt did change the correspondence between micrometer and CPS readings. Alignment between masses was established in room temperature at atmospheric pressure. At low temperatures, with the system under vacuum, the capacitive sensors and the room temperature calibrations (rather than the micrometer readings) were used to indicate the relative position and tilt between stages. 

\subsection{Cryogenic Apparatus}
Cooling the cantilever was crucial to achieving a high force sensitivity. The probe was sealed in a vacuum can inside a $^4$He cryostat, with an exchange gas space separating the probe from the mechanical vibration of the boiling helium. The three-axis positioner was located outside of the cryostat, attaching to the bimorph stage via a vacuum feedthrough. The interferometer was also located outside the cryostat, with a length of optical fiber running down the probe. 

Base temperature of the probe was typically 10 K, though this temperature was raised a few degrees by the use of spring-adjustment heaters (described in Sec.~\ref{alignmentsec}). The noise temperature of the cantilever was found to be typically $\sim$~10~K more than the base temperature of the probe. 

Hold time for the cryostat was typically 4--6 days. After each helium transfer, all preliminary tests of the system were performed. Each data set presented in this paper was recorded entirely within one helium transfer of the respective cooldown. 

\subsection{Interferometer}\label{intfsec}
Cantilever motion was measured with a fiber interferometer, based on the design described in Ref.~\cite{rugarintf}. An InGaAs laser diode sent hundreds of microwatts of 1310-nm light through a bidirectional fiber coupler, leaving $1\%$ of this light for the length of fiber that went to the cantilever. The cleaved end of the fiber in the probe was aligned to the test mass, forming a low-visibility Fabry-Perot cavity with a length of $\sim 50$~$\mu$m. Interference between the light reflected off the test mass and the light reflected from the cleaved end of the fiber was measured with a photodiode via a transimpedance amplifier with a 10~M$\Omega$ feedback resistor. 

Because of the low reflectivity of the cleaved end of the optical fiber, beam divergence, and imperfect alignment between the fiber and the test mass, only one reflection from each surface (the fiber end and the test mass) was expected to contribute to the interferometer signal. Thus, the dc interferometer signal was a sinusoidal function of the distance between the fiber end and the test mass, modulo half the wavelength of the laser. The sensitivity of the interferometer was maximized when the distance between the cantilever and the optical fiber was adjusted to be at one of the points of maximum slope of this sinusoidal fringe (the center of the fringe). 

Cantilever motion was typically on the order of angstroms at low temperatures. For motion much less than the wavelength of the laser, the interferometer signal at a given frequency $V_i(f)$ was linearly related to the amplitude of cantilever motion at that frequency $x(f)$. The conversion between the voltage signal from the interferometer and cantilever motion depended on the peak-to-peak amplitude $V_{pp}$ of the sinusoidal interferometer fringe and the relative position between the cantilever and the optical fiber that determined the location on this fringe of the interferometer dc level. When the distance between the fiber and the cantilever was adjusted to the center of the fringe, the amplitude of cantilever motion was determined by \begin{equation}\label{vppeqn}x(f)=V_i(f) \frac{\lambda_l}{2\pi V_{pp}} ,\end{equation}
where $\lambda_l$ is the wavelength of the laser.  

Temperature control and high-frequency ($> 100$~MHz) modulation \cite{rugar_ultramic} were used to stabilize the laser during data acquisition. The high-frequency modulation shortened the coherence length of the laser, reducing the importance of stray reflections from connectors within the optical part of the interferometer circuit.

\subsection{Cantilever Position Adjustment and Characterization}
A piezoelectric stack (hereafter referred to as ``piezo stack'') mounted underneath the cantilever wafer was used to adjust the distance between the cantilever wafer and the optical fiber to maintain alignment at the center of the interferometer fringe. This piezo stack was also used to excite the cantilever for the purposes of characterization. 

In the probe, the piezo stack was mounted between the wafer frame and the wafer stage. Upon cooling, the stainless steel wafer frame and stage would contract while the piezo stack would lengthen slightly.  At low temperatures, it was expected that the differential thermal contraction of the materials would have made the wafer frame tilt in the $x$-$z$ plane. A correction for this tilt was included in the room temperature alignment between masses. 

\section{Finite Element Analysis and Spatial Phase-Sensitive Detection}
In order to deduce a bound on Yukawa-type deviations from Newtonian gravity, the measurement must be compared to the expected force between the masses for both a Newtonian potential and a Yukawa potential. Finite element analysis (FEA) with a mesh size of 5~$\mu$m was used to calculate the expected dc (without time variation) gravitational force between the masses for a range of longitudinal (along the $y$-axis) positions between the masses, with the vertical separation held constant. 

At a given $y$-point (with other alignment parameters set), the dc Newtonian force $F_N^{\rm (dc)}$ and dc Yukawa force  $F_Y^{\rm (dc)}$ (with $\alpha=1$) were calculated by sums over the drive mass and the test mass in the 5~$\mu$m mesh. Only the vertical component of the force was considered and the attractive force was taken to have a positive sign. The Newtonian force sum is given by:
\begin{equation}\label{FEAnewton}
F_N^{\rm (dc)}=\sum_{V_{\rm d}}\sum_{V_{\rm t}}\left[\left(\frac{G\delta m_{\rm d}\delta m_{\rm t}}{r^2}\right) \left(\frac{z}{r}\right)\right].
\end{equation}
\noindent Here, the two sums are over the volumes of the drive mass and the test mass in the given mesh, $\delta m_{\rm d}$ refers to the mass of the (5~$\mu$m)$^3$ block in the drive mass, $\delta m_{\rm t}$ refers to the mass of the (5~$\mu$m)$^3$ block in the test mass, and $r$ is the distance between the centers of mass of these two blocks in the summation. The vertical separation between the two mass blocks is $z$; the final term in the equation is a projection onto the vertical axis. Similarly, the Yukawa force (for $\alpha=1$) for a given value of $\lambda$ was calculated from the sum:
\begin{equation}\label{FEAyuk}
F_Y^{\rm (dc)}=\sum_{V_{\rm d}}\sum_{V_{\rm t}} \left[\left(\frac{G \delta m_{\rm d} \delta m_{\rm t}}{r^2}\right) e^{-r/\lambda}(1+r/\lambda)\left(\frac{z}{r}\right)\right].
\end{equation}

A time-variation (accounting for drive mass motion) was applied to this calculated set of dc forces as a function of $y$-position and from this, the expected ac (finite frequency) force was extracted with a Fourier transform, yielding ac Newtonian and Yukawa forces. In the case of Cooldown~04, the third harmonic of the drive frequency was studied; the Fourier component of the measured force at this frequency is referred to as the third harmonic ac force.

There were several inputs to the FEA model: the geometry and density of the masses, the ($x$, $y$, $z$) position and tilts between masses, amplitude of bimorph motion, and the range $\lambda$ of the Yukawa potential being modeled. The output of the model was the ac Newtonian gravitational force between masses $F_N$ and the ac Yukawa force $F_Y$ for $\alpha=1$ and a given $\lambda$. For arbitrary $\alpha$, the Yukawa force could simply be scaled by $\alpha$. 

\begin{figure}
\includegraphics[width=\columnwidth]{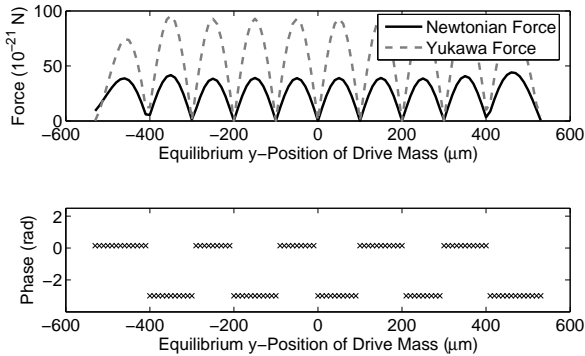}
\caption{The calculated magnitude (top) and phase (bottom) of the predicted third harmonic ac gravitational force between masses. The larger Yukawa force for $\alpha=5$ and $\lambda=34$~$\mu$m is shown by the dashed line; the smaller Newtonian force is the solid line. The phase is the same for a Newtonian or a Yukawa force. In this calculation, the drive mass was taken to be  27~$\mu$m underneath the test mass, without tilt. The middle gold bar of the drive mass is centered under the test mass at $y$-position of 0~$\mu$m.\label{acforcephase}}
\end{figure}

Both the magnitude and phase of this Fourier component with respect to the drive frequency are important. Clearly, the dc gravitational force between the masses reflects the periodicity of the drive mass pattern. In fact, the ac force also shows this periodicity. As shown in Fig.~\ref{acforcephase}, the magnitude of any ac gravitational force has a clear periodicity of 100~$\mu$m as a function of the $y$-equilibrium position of the drive mass with respect to the test mass, corresponding to the 100~$\mu$m half-period of the drive mass pattern. Each minimum of the third harmonic ac force magnitude is zero and is accompanied by a discontinuous phase change of $\pi$. The fourth harmonic ac force (not shown) has the same periodicity, though the fourth harmonic force has minima where the third harmonic force has maxima, and vice-versa. 

To exploit this geometric feature of our design, data were recorded at several values of the $y$-equilibrium position between masses, scanning over more than 200~$\mu$m, the period of the drive mass pattern. Comparison of these $y$-scan measurements to FEA predictions allowed us to discriminate between couplings that could be gravitational in origin and spurious backgrounds that do not follow the gravitational pattern. By ``locking-in'' this way to the expected spatial periodicity, we were able to set a stronger and more accurate bound on $\alpha(\lambda)$ than what a single force measurement would have provided. 

\section{Experimental Methods}
\begin{figure}
\includegraphics[width=\columnwidth]{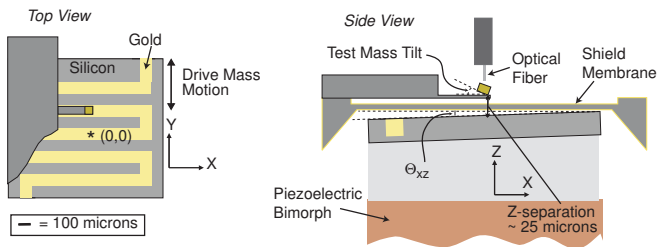}
\caption{[color online] Schematic showing the geometry and alignment parameters between the test and drive masses. On the left, the top view shows the $x$-$y$ plane; the origin, marked with an asterisk, is the center of the drive mass in this plane. With the drive mass at equilibrium, the $x$-$y$ position of the test mass was defined with respect to this origin. This figure shows the masses to scale; the shield membrane and the ground plane over the drive mass are omitted. On the right, the side view shows (not to scale) the $z$-$x$ plane. The $z$-separation between masses is the face-to-face vertical distance. The test mass tilt, exaggerated in this figure for the purposes of illustration, is the tilt of the test mass with respect to the cantilever. The angle between the drive mass and the cantilever in this plane was defined as $\theta_{xz}$.\label{alignschem}}
\end{figure}

In order to accurately compare measurements to the FEA, alignment between masses had to be known as accurately and precisely as possible. In order to maximize the gravitational force between masses, the goal was to center the masses with respect to each other with no tilts about the $y$-axis ($\theta_{xz}$) or the $x$-axis ($\theta_{yz}$). The alignment coordinates are diagrammed in Fig.~\ref{alignschem}. With no tilts in the system, the third harmonic ac gravitational force would be maximized with the test mass at $x=0$~$\mu$m and $y=50$~$\mu$m with respect to the center of the drive mass at equilibrium. 

Alignment between the masses was fixed by room-temperature preparations and low-temperature adjustments. This process began with the assembly of the cantilever and shield wafers.

\subsection{Wafer Assembly}
The test mass was attached to the cantilever using epoxy applied with a microprobe under an optical microscope. The optical microscope used for this process allowed selection of the better test masses from the fabrication batch.

After attaching test masses to cantilevers, the cantilever wafer was glued into a stainless steel wafer frame. The shield wafer was then glued to this cantilever wafer, using a press to keep the wafers parallel. Measurements before and after each gluing showed the degree of tilt ($\ll 1$~mrad) between the two wafers. Photographs of the two wafers before and after gluing were used to determine the relative position between each cantilever on the wafer and the center of the shield membrane. 

By design, the shield membrane was 10~$\mu$m below the surface of the shield wafer that was glued to the cantilever wafer. The glue between the wafers added typically 1--5~$\mu$m to this distance. The offset of the shield and the thickness of the glue layer did limit the minimum vertical separation between masses. However, reducing this distance significantly would cause the cantilever to snap-in and adhere to the shield \cite{stiction}.

\subsection{Probe Assembly}
Before mounting in the probe, the shield membrane was carefully examined for dust. Any pieces of dust with a resolvable height profile ($\ge 2$~$\mu$m) were removed with a microprobe. Then the wafer was mounted in the wafer stage and the optical fiber was aligned to the test mass. Calibrations of position sensors were performed and the drive mass and the bimorph were mounted in the probe. 

\subsection{Drive Mass Gluing}
The drive mass was glued to the bimorph with the bimorph positioned so that the drive mass was pushed against the silicon shoulder of a shield wafer, mounted in the probe. This process, only repeated when the bimorph or the drive mass needed to be replaced, achieved approximate parallelism between the drive mass and any cantilever wafer. Before each cooldown, parallelism between the shield wafer and the drive mass was again examined and adjusted. Uncertainties in the tilt between the drive mass and the cantilever were dominated by uncertainties in this process. 

\subsection{Alignment}
After the drive mass was glued to the bimorph, the parallelism between the shield wafer and the drive mass was examined optically, using images recorded from a telescope via a CCD camera. Parallelism was assessed in the $x$-$z$ and $y$-$z$ planes by looking at the reflection of the drive mass in the gold-coated shoulder of the shield wafer. CPS readings and optical images were compared as the bimorph stage was moved to change the vertical separation between the drive mass and the shield wafer. Tilts measured in two vertical planes ($\theta_{xz}$ and $\theta_{yz}$) were adjusted using turnbuckles on the lower vibration isolation stage (the wafer stage), with additional compensation for the expected piezo-induced tilt of the wafer frame upon cooling. Uncertainty in this part of the alignment was dominated by optical limitations. 

The large three-axis positioner was used to adjust the alignment of the drive mass with respect to the test mass both at room temperature and at low temperatures. Micrometers on this stage were accurate to 2.54~$\mu$m (0.0001~in), with the $y$ and $z$ micrometers controlled by motors.  

After optimization of the tilt, the bimorph frame was moved so that the drive mass was centered in the $x$-$y$ plane with respect to the shield membrane; a telescope with a reticle aided this centering process. The bimorph stage was then slowly raised until the drive mass contacted the shield. This contact gave an impulse to the shield and the cantilever, creating a clear signal on the interferometer. CPS and tilt sensor readings were recorded several times during this process; these alignment points were the targets for initial alignment when the system was cold. Uncertainties in this process included optical limitations and motion of the three-axis positioning stage in the $x$-$y$ plane as it was raised along the $z$-axis.

The drive mass and shield were examined through a telescope for any indication of dust on either surface before the vacuum can was closed. 

\subsection{Re-Alignment\label{alignmentsec}}
Under vacuum, the tilt of the positioning stage connected to the bimorph frame significantly changed the alignment between the wafer stage and bimorph stage, and hence the alignment between masses. At base temperature, the tilt of the wafer stage was adjusted by heating two of the three springs on the upper vibration isolation stage. To heat a spring, a current on the order of 20~mA  was passed through a manganin wire (of resistance $\sim 50$~$\Omega$) wrapped around the coils of the BeCu spring; such heating could increase the length of the spring on the order of 100~$\mu$m. Temperature and position of the probe stabilized after several hours. 

The long time scale of the spring heating prohibited fine adjustment of the tilt. However, the large separation between the tilt sensors ($\sim$~38~mm) in the $x$-$y$ plane in comparison to the size of the drive mass ($\sim$~1.5~mm) allowed for a coarse readjustment of tilt at low temperatures; typically, the vertical separation between the tilt sensors for a given CPS $z_c$-reading was within 100~$\mu$m of the room temperature alignment points.

With the tilt adjusted, the $x$-$y$ position of the bimorph stage was adjusted to regain the room temperature alignment points. The manual operation of the $x$-micrometer necessitated coarser alignment in this direction, since the micrometer could not safely be adjusted with the drive mass positioned close to the shield wafer. However, the alignment between the masses in the $x$-direction only needed to be accurate to within a couple of hundred microns, since the Yukawa forces being studied were short range in comparison to the 1-mm length of the drive mass bars. The motor control of the $y$ and $z$ micrometers allowed adjustments in these directions to within a micron, with uncertainty being dominated by noise (electrical and mechanical) on the capacitive readings.

Additional information about the alignment between masses was provided by moving the bimorph frame in the $y$-direction until the drive mass contacted the side of the silicon frame bearing the shield membrane. This test was performed with the bimorph static and lowered so that the drive mass was $\sim 200$~$\mu$m away from the shield. The contact between the drive mass and the wafer gave a mechanical impulse to the cantilever, clearly visible on the interferometer signal. The distance between the alignment point and this contact point provided additional confirmation of the drive mass alignment along the $y$-axis. 

After realignment, the bimorph frame could then be moved to the $y$-position, determined by the photographs of the wafer, at which the drive mass was centered underneath the test mass. The experimental values and uncertainties in the alignment parameters are given in Tbl.~\ref{alignmenterror}.

        \begin{table}[t]
        \begin{center}
        \begin{minipage}{\columnwidth}
        \begin{center}
        \caption{Alignment between drive mass and test mass }
        \smallskip
        \begin{tabular} { l  r  r  r}
        \hline
        \hline
        Parameter & Value & Error &Units\\
        \hline		
         $x$-position	& -91	&110	&$\mu$m\\
         $y$-position &(-12)--(283)	&119\footnote{This is the error on the location of the entire range. Within this range, there was an uncertainty of about 5 $\mu$m on each point.}	&$\mu$m\\
         $z$-separation 	&24--28	&3.4&$\mu$m\\
	$|\theta_{xz}| \times$ width\footnote{Here, width is of the drive mass die: 1.3 mm.} 	&1	&6	&$\mu$m\\
	$|\theta_{yz}| \times$ length\footnote{Here, length is of the drive mass die: 1.8 mm.}	&6	&9	&$\mu$m\\
        \hline
        \hline
        \end{tabular}
        \label{alignmenterror}
        \end{center}
        \end{minipage}
        \end{center}
        \end{table}

\subsection{Vertical Positioning of Drive Mass}
Typically, gravitational measurements were recorded with the drive mass positioned 10--15~$\mu$m away from the shield, with this safety factor allowing for the small vertical motion of the drive mass over the course of the bimorph swing, backlash in the motor used to operate the $z$-axis of the three-axis positioner, and the possibility of drift during a data run.  

To determine the vertical ($z$-) separation between the masses, the bimorph frame was slowly raised until the drive mass contacted the shield, giving an impulse to the cantilever clearly visible on the interferometer. The bimorph was then lowered to the desired distance from the shield with the CPS indicating the amount of motion. The largest sources of uncertainty in the $z$-separation between masses were bouncing during contact and data acquisition and the possibility of dust on the shield or drive mass. Due to lack of precision in setting a given $z$-separation (due to limitations of the motor driving the $z$-micrometer), the $z$-separation during data acquisition varied over the course of a day; this is the reason for the range given in Tbl.~\ref{alignmenterror}.  

\subsection{Cantilever Characterization}
After reaching base temperature, the piezo stack was used to verify the interferometer fringe. The shape of the fringe was an important indicator of alignment between the optical fiber and the test mass. In earlier designs of this apparatus, the fiber alignment often drifted upon cooling so that more light was reflecting from the shield than from the test mass. Other times, light reflected from the cantilever as well as from the test mass. The former situation created a large background level on the interferometer; the latter would make it impossible to record data. In Cooldown~04, the shape of the fringe indicated no misalignment between the optical fiber and the test mass at low temperatures. 

The resonant frequency $f_0$ was determined by exciting the cantilever at large amplitudes with the piezo stack and comparing the amplitude and phase of the resulting signal to the drive signal. The quality factor was determined by timing the ringdown of the cantilever from excitation on resonance. The fringe height, the resonant frequency, the quality factor, and the dc level of the interferometer on the fringe were checked between data collection runs. 

\subsection{Bimorph Actuation}
After alignment between the masses was established, the bimorph was lowered to a safe distance and an ac voltage was applied to the bimorph via a high voltage amplifier. The capacitive measurement of the bimorph motion indicated that the bimorph motion for a given driving voltage increased over time, typically coming to equilibrium within 30 min. 

\subsection{Experimental Degrees of Freedom}
To make the most sensitive gravitational measurement, data were recorded as a function of $y$-position between masses, with a small vertical separation between masses and the drive frequency $f_d$ tuned to be $f_0/3$. Diagnostics included data runs with $f_d$ tuned slightly off this resonance, with the drive mass far away from the test mass, with the bimorph not moving, or with $f_d=f_0/4$. 

\section{Data Acquisition and Averaging}
Two streams of time-series data were recorded via an analog-to-digital converter (ADC) on a personal computer: the voltage from the interferometer and the voltage from the function generator that was driving (through a high-voltage amplifier) the bimorph. The interferometer signal showed the cantilever motion. The function generator signal (at frequency $f_d$ close to 100~Hz) provided an important timing signal. Data were recorded at a frequency of 10 kHz. Before the ADC, the interferometer signal was ac-coupled to a pre-amplifier with a high-frequency rolloff at 3 kHz to avoid aliasing. 

Any motion of the cantilever that was due to the moving drive mass would be at a definite phase with respect to the drive mass motion and at a harmonic of the drive frequency. The function generator signal  (the drive signal) provided a proxy for the drive mass motion and the analysis of the cantilever motion then included a phase defined with respect to this drive signal.

Each data run included data recorded over a period of time $t_{t}$, with time-series data examined in shorter records of time $t_0$. The time $t_0$, typically on the order of $Q/f_0$, was chosen so that each record of length $t_0$ could be considered statistically independent of other records, while maximizing the total number of records $N=t_{t}/t_0$.

For each record of length $t_0$, the time-series data were truncated to include an integer number of periods of the function generator signal. This truncation shortened the length of each record a negligible amount and increased the accuracy of the amplitude and phase of the harmonics of $f_d$ reported by a Fourier-transform of the record.

The drive signal, resulting from a very clean function generator signal, showed an unambiguous peak on the Fourier transform at the drive frequency $f_d$. The harmonic of interest of this drive frequency was then selected from the Fourier-transform of the interferometer signal. In the case of Cooldown 04, measurements were made from the third harmonic of the drive signal, intended to be equal to $f_0$. The signal from the interferometer was converted to an amplitude of cantilever motion, as described in Eq.~\ref{vppeqn}. From this, the third harmonic ac force on the cantilever (assuming the $3f_{d}=f_0$) was determined using Eq.~\ref{acforceeqn}.

The magnitude and phase (or, equivalently, the real and imaginary parts) of the Fourier transform are important to consider. For each record $t_0$, the real $F_R$ and imaginary $F_I$ parts of the Fourier transform at $3f_d$ were found, providing $N$ measurements of $F_R$ and $F_I$. Averaging each of these two sets provided the means $\overline{F_R}$ and $\overline{F_I}$ for the given data run, with a standard error on each mean determined by the standard deviation of the set, reduced by a factor of $\sqrt{N-1}$. 

Aside from the statistical error, uncertainty in the assessment of the force on the cantilever from this averaging technique resulted from uncertainty in the fringe height of the interferometer, the interferometer dc position relative to the center of the fringe, the cantilever $Q$, and the setting of $f_d$ to be exactly $f_0/3$. Furthermore, Eq.~\ref{acforceeqn} assumes that the measured displacement of the cantilever is at the center of mass of the test mass. Due to the mode shape of the cantilever, described analytically in Ref.~\cite{laura} and modeled in ANSYS, the deflection at either end of the test mass would be different by $\sim$~15--20$\%$ from the deflection at the center of mass. Though the quality of the interferometer signal indicated that the optical fiber was focused on the test mass (instead of the cantilever), the degree of freedom in the fiber position along the length of the test mass provided another uncertainty in the conversion between interferometer signal and force. These uncertainties are summarized in Tbl.~\ref{forceerror}.

    \begin{table}[t]
        \begin{center}\begin{minipage}{\columnwidth}
        \begin{center}
        \caption{Uncertainties in measured force magnitude}
        \smallskip
        \begin{tabular} { l  r  r  r r}
        \hline
        \hline	
        Parameter & Value\footnote{Values given are typical for Cooldown~04, with feedback cooling on. For parameters (such a fringe height) that varied over the course of a day, the error given does not include long time scale drift, only the uncertainty over the course of an hour-long data run.} & Error &Units &$dF$\footnote{This is the proportional error in the force measurement resulting from the error in the named parameter.} ($\%$)\\
        \hline
	Fringe height		&0.228		&0.002		&V		&1 \\
	Fringe position from center		&0			&10			&$\%$	&2\\
	$f_0$\footnote{The amplitude of the Lorentzian transfer function is extremely nonlinear as a function of frequency. Because the error considered here makes an assumption of linearity, the error in $f_0$ is overestimated to account for the nonlinearity; the relative error in the force was taken to be half of the change in the measured force that would result from twice the experimental uncertainty in the frequency.}
									&331.178		&0.002	&Hz 		&2\\
	Q								&10000	&$1000$		&-- 		&10\\	
	$k$								&0.0071		&0.001		&N/m	&8	\\	
	Fiber alignment 					&0			&25			&$\mu$m	&10	\\
         \hline
         \textbf{\textit{Total}}				&			&			&$\%$N		&17\\      
         \hline
        \hline
        \end{tabular}
        \label{forceerror}
        \end{center}
        \end{minipage}\end{center}
        \end{table}

\subsection{Feedback Cooling}
A high quality factor was essential to achieving a sensitive force measurement. However, a high quality factor also made measurement difficult; the resonant frequency had to be determined to an accuracy of $\sim f_0/(3Q)$ and between data runs or after any excitation of the cantilever, a waiting period of at least $\sim 3Q/f_0$ had to be observed to allow the cantilever to ringdown. In Cooldown 04, $Q\sim 80000$. To facilitate the data-acquisition process while maintaining the low thermal noise of the high quality factor cantilever, feedback cooling was used to reduce the effective quality factor and the noise temperature $T$ of the cantilever in its lowest mode. 

\begin{figure}
\includegraphics[width=\columnwidth]{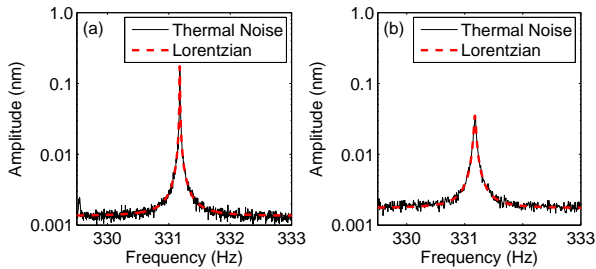}
\caption{[color online] Measured thermal noise data without (a) and with (b) feedback are shown in comparison to the Lorentzian curves derived from the measured $k$, $Q$, effective $T$, and $f_0$. The rms amplitude of cantilever motion is plotted as a function of frequency. The offsets added to the Lorentzian curves are judged from the power spectrum. The background in the figure on the right is higher because of the larger bandwidth. With feedback, $Q\sim$~10400 and the spectrum is the result of averaging thirty power spectra from 2 min data records; the effective temperature was $\sim$~3.5~K. Without feedback, $Q\sim$~75000 and the spectrum is an average of thirty records of length 4~min each; the effective temperature was $\sim$~30~K. \label{lorentzianfit}}
\end{figure}

The feedback-cooling apparatus was based on the designs of the Rugar group \cite{rugar_pc} and the feedback circuit described in Ref.~\cite{jacthesis}. The interferometer signal was phase-shifted, attenuated, and used to drive the piezo stack. In this way, the cantilever was driven with a phase-shifted version of its own thermal noise. The circuit was adjusted to incorporate a phase shift of $\pi$ over the thermal noise bandwidth of the first mode of the cantilever. Over this small bandwidth, this additional phase shift turned the feedback into negative feedback on the velocity (rather than on the position) of the cantilever. A cantilever is a damped simple harmonic oscillator; negative feedback on its velocity increases the damping, decreasing $Q$. It also decreases the amplitude of all thermal noise excitation in this first mode in a way that decreases the effective temperature of the cantilever in that mode.  The thermal noise spectra of the cantilever in Cooldown~04 with and without feedback are shown in Fig.~\ref{lorentzianfit}.

Within the accuracy of the measurement of these two parameters, the ratio of $Q/T$ was maintained with this feedback and thus the low thermal noise limit was not affected by the reduction of the quality factor. However, the sensitivity of the cantilever, as described by Eq.~\ref{acforceeqn}, does depend on $Q$; feedback cooling in Cooldown~04 did reduce the voltage signal-to-noise ratio of the measurement.

\section{Tests of the Apparatus}\label{testssec}
\subsection{Thermal Noise}
To verify the functioning of the cantilever, to confirm the conversion from interferometer signal to amplitude of cantilever motion, and to determine the force limit in the experiment, thermal noise of the cantilever was recorded. During thermal noise measurements, the bimorph was turned off. Thermal noise may be examined several ways. The noise temperature of the cantilever was determined from power spectra averaged over several consecutive time records. The equipartition theorem relates the mean squared displacement $\left<x^2\right>$ in one mode of the cantilever with spring constant $k$ to the temperature $T$: \begin{equation} \left<x^2\right>=k_B T/k .\end{equation} By summing the squared amplitudes of cantilever motion over the thermal noise peak, the effective temperature was found. The white noise background of the interferometer was subtracted from this summation and the sum only included the thermal noise bandwidth so as to exclude electronic noise on the interferometer. The power spectrum of the cantilever due to thermal noise compared well to the expected Lorentzian, as shown in Fig.~\ref{lorentzianfit}. Though the temperature of the cantilever without feedback was found to be $\sim 30$~K, higher than the temperature of the probe, the agreement between the measured noise spectrum and the Lorentzian function using the measured values of $Q$, $k$, and $T$ confirmed our assessment of these important parameters. 

Thermal noise is random phase noise. Though a cantilever is constantly excited by the energy in its thermal environment, this excitation comes as a series of random-phase kicks, with the ringdown from each kick characterized by the quality factor. When the real and imaginary parts of a given Fourier component of thermal noise are considered, the measurements (over many time records) show a force that is statistically indistinguishable from zero. In this case, the mean of the set of measurements of $F_R$ should be less than twice the standard error on the mean: $\overline{F_R}<2\sigma_R/\sqrt{N-1}$, where $\sigma_R$ is the standard deviation in the set of $N$ measurements. The same is true for the set of measurements of $F_I$ or any other phase of any Fourier component. The measured force $(\overline{F_R}^2+\overline{F_I}^2)^{1/2}$ due to thermal noise will decrease with the square root of the averaging time, as seen in Eq.~\ref{thnoiseeqn}.

Even in the case of a coherent driving force on the cantilever, where the measured force at some phase is statistically distinguishable from zero, thermal noise is important to consider---thermal noise provides the statistical uncertainty on any force measurement. Averaging longer reduces the statistical uncertainty due to thermal noise. 

\subsection{Magnetic Analog Experiment}
In Cooldown 02, the buried drive mass and a magnetic test mass were used in order to test the couplings between masses via a large, measurable magnetic force. The probe design was an earlier version than the one described above (used in Cooldown 04) and alignment methods were less precise. Nonetheless, this experiment served as an important demonstration of the functioning of the apparatus. 

To couple the drive mass to the magnetic test mass, an ac voltage could be applied to the drive mass with the bimorph not moving or a dc current could be drawn through the drive mass with the bimorph moving. In either case, the magnetic coupling between the drive mass and the test mass would reflect the 200~$\mu$m half-periodicity of the magnetic field across the drive mass. The exact direction of the dipole moment of the test mass, not known, would strongly influence both the magnitude of the coupling as well as the exact dependence of the force magnitude on the $y$-position between the masses. However, FEA showed that any magnetic force would have a distinct periodicity of 200~$\mu$m in the magnitude of the force, with each force minimum accompanied by a phase change of $\pi$. These features are qualitatively similar to those of the gravitational force shown in Fig.~\ref{acforcephase}, though the spatial period across the drive mass of this magnetic coupling would be twice that of a gravitational coupling.

To quickly confirm the position of the drive mass with respect to the test mass, 0.5 V ($\sim 2$~mA) at $f_0$ was applied across the drive mass meander, with the bimorph not moving. With the drive mass close to the test mass, this created a large excitation of the cantilever, clearly visible on the interferometer signal viewed on an oscilloscope. The drive mass was moved underneath the test mass and from this, the locations (in the CPS coordinates) of the distinct phase changes in the magnetic force were noted and used as reference in the measurements that followed. 

\begin{figure}
\includegraphics[width=\columnwidth]{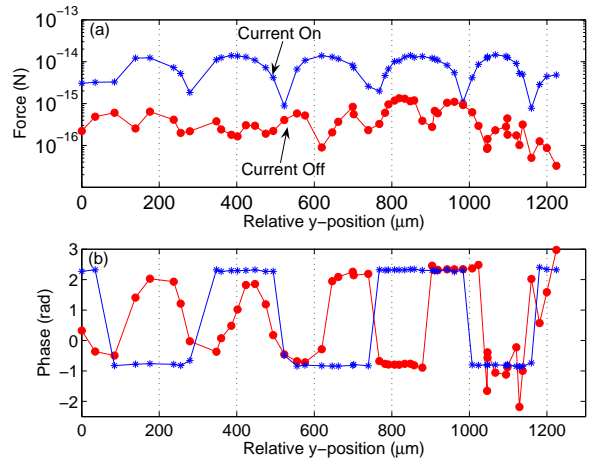}
\caption{[color online] The third harmonic force magnitude (a) and phase (b) as a function of $y$-position (relative to the first point) between the magnetic test mass and the drive mass with current on and off. The magnitude here is calculated from the mean of the real and imaginary components: $\sqrt{\overline{F_R}^2+\overline{F_I}^2}$.\label{bigmagyscan}}
\end{figure}

A $y$-scan was performed with the bimorph moving $\sim$~100~$\mu$m in amplitude at $f_0/3$. Measurements were made at intervals of $\sim 25$~$\mu$m in $y$, covering the entire drive mass pattern. At each $y$-position, measurements were made both with a current (0.5 mA) across the drive mass and with no drive mass current. Even with short averaging times (20 s with a current on and 10 min with no current), a clear force was measured in each case. The results of this $y$-scan are shown in Fig.~\ref{bigmagyscan}.

With the current drawn across the drive mass, the force magnitude showed five distinct periods, corresponding to the five sets of gold and silicon bars. As expected, the phase of the force as a function of $y$-position showed distinct discontinuous phase changes of $\pi$ associated with each force minimum. Without knowing the direction and size of the magnetic dipole moment of the test mass, this measurement could not be used as a precise calibration. Furthermore, a broken shield prevented a precise assessment of the vertical separation between masses. Nonetheless, this measurement provided a confirmation of the alignment between masses, the FEA predictions, and the relative position between data points as determined by the CPS. 

Measurements with no current across the drive mass also showed a force with a distinct periodicity, though in this case the periodicity was 100~$\mu$m in the magnitude of the force, as expected for a gravitational force. Phase changes of $\pi$ were also evident in this measurement, though the phase changes were continuous. Continuous phase changes would be expected if there was a constant force in addition to the varying force across the $y$-scan. 

The measured force in the case of no drive mass current was much larger than any expected gravitational force \cite{frogs}. The shield and the buried drive mass would not permit any electrostatic or Casimir force to drive the cantilever in a way that would show this periodicity of the drive mass. This force could only be magnetic in origin. 

Silicon and gold have small diamagnetic susceptibilities, on the order of $-0.16 \times 10^{-8}$~m$^3$/kg. Due to the differing mass densities of the gold and the silicon, each bar of silicon has a susceptibility that is smaller than the gold bar of the same size. Magnetized in any ambient field, the drive mass will have a spatially-varying magnetic field even in the absence of applied current.  This varying magnetic field will couple to the test mass and in this way, the magnetic test mass will act as a susceptometer. Though this effect is small, our experiment is sensitive enough to measure such a coupling; a rough calculation shows that even in the earth's magnetic field, this effect can be as large as the measured force shown in Fig.~\ref{bigmagyscan}. 

The functioning of the magnetic test mass as a susceptometer limited the sensitivity of any gravitational force measurement with the magnetic test mass. However, the measurement of both forces with the magnetic test mass---including both the 200~$\mu$m and the 100~$\mu$m periodicities in the force magnitude---provided an excellent verification of the sensitivity of the cantilever as a force detector and demonstrated that the experimental apparatus can measure the coupling between masses.  

\subsection{Possible Background Couplings}
With a non-magnetic test mass, the buried drive mass and the shield eliminated the possibility of magnetic, electrostatic, or Casimir couplings to the test mass that could be mistaken for gravity. Couplings due to induced currents in both masses, magnetic impurities in the drive mass, or induced moments in both masses were all estimated to be much less than the target thermal noise limit of 10$^{-18}$~N. The possibility of a coherent gravitational excitation of the cantilever due to anything other than the drive mass was made unlikely by the extremely high quality factor of the cantilever and the relatively high resonant frequency as compared to the motion of commonplace objects or people in the laboratory. 

If the vibration isolation was shorted, the bimorph could have shaken the cantilever. Measuring with the third harmonic $3f_d$ on- and off-resonance gave an indication of whether the vibration isolation was failing. Cantilever motion due to a mechanical excitation would be reduced by a factor of $Q$ off-resonance. 

Finally, there could have been electrical or mechanical coupling between the bimorph and the interferometer that was unrelated to the cantilever. Such electrical or mechanical couplings to the interferometer would not vary as a function of the drive mass equilibrium $y$-position and thus would never be mistakenly interpreted as a gravitational signal. However, the resulting voltage noise could impede the detection of a small gravitational signal.

Electrical coupling could result from imperfect grounding of the function generator or the circuitry of the laser or the interferometer. Mechanical coupling could have resulted from the optical fiber being shaken by the moving bimorph. In both cases, the coupling would produce a voltage noise on the interferometer that would be insensitive to small changes in frequency; the signal on the interferometer (with the subtraction of thermal noise) would be the same on- and off-resonance. Examinations of signals at $f_d$ and $2f_d$ also provided a helpful diagnostic of such couplings.

The optical fiber ran down the length of the tube to which the bimorph frame was attached. Motion of the bimorph could couple to the optical fiber itself, shaking the fiber at $f_d$. Nonlinearities in the bimorph itself and in any part of the mechanical path between the bimorph and the optical fiber could result in the fiber also being shaken at harmonics of $f_d$. Because there were reflections within the optical part of the interferometer circuit (due to the finite reflectivity of all fiber connectors) that created stray interferometric paths, vibration of the optical fiber could create a measurable signal on the interferometer at the frequency of vibration. Modulating the laser at high frequencies to reduce the coherence length could reduce but not eliminate this kind of noise. Having a large fringe height and a high quality factor to increase the sensitivity of the interferometer as described by Eq.~\ref{acforceeqn} and Eq.~\ref{vppeqn} would reduce the importance of this kind of background noise.

\section{Experimental Results}
In Cooldown 04, a sensitive gravitational measurement was made using a nonmagnetic test mass, a buried drive mass, a new probe design to improve fiber alignment to the test mass, and more precise alignment techniques between the masses. A $y$-scan covering almost 300~$\mu$m was recorded over two days. On each day, measurements were made at six points. The first data run on Day~1 was 48 min long; all other data runs were 60 min in length. Data were recorded in records of 4 min in length, though analysis studied the records in 30 s segments. Feedback cooling was used to reduce the $Q$ of the cantilever from $\sim$~80000 to $\sim$~10000. The bimorph was oscillated at an amplitude (at the top surface of the drive mass) of 125~$\mu$m. 

\begin{figure}
\includegraphics[width=\columnwidth]{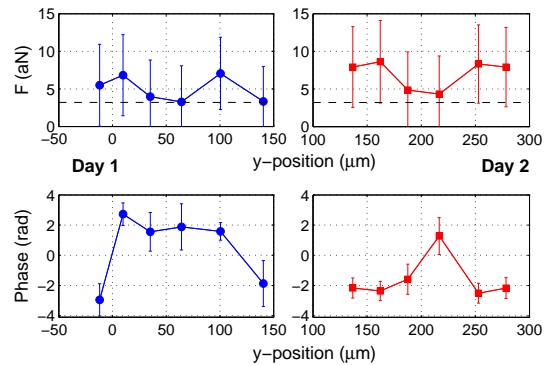}
\caption{[color online] Force magnitude (top) and phase (bottom) vs.\ $y$-position from Cooldown~04. Day~1 data are on the left and Day~2 data are on the right. The black dashed line shows the approximate level of measured thermal noise, which was slightly less than the predicted thermal noise level. Plotted points are the means for the full averaging time (60 min, except for the first measurement, which was 48 min); the error bars show twice the standard (statistical) error on the mean. Lines connecting the measured points are guides to the eye.\label{results}}
\end{figure}

The measured force and phase for the $y$-scan are shown in Fig.~\ref{results}, along with the thermal noise limit. On both days, the measured magnitude of the force showed an apparent periodicity of 100~$\mu$m, though there was no clear periodicity in the measured phase. All measured forces were close to thermal noise in magnitude. 

On Day~2, the measured force was consistently higher than on Day~1. The interferometer signals at $f_d$ and $2f_d$ were also larger on Day~2, suggesting the presence of some mechanical or electrical coupling to the interferometer. 

The error bars in Fig.~\ref{results} show twice the statistical error on the measurements; the local maxima in the force measurements are statistically distinguishable from zero. 

\section{Analysis}
The measurements from Cooldown 04 do not clearly look like a gravitational force. Even if part of the measured force was gravitational in origin, some of the measured force was certainly due to thermal noise and possibly due to some other background. To set a bound on Yukawa-type deviations from Newtonian gravity, these measurements as a function of $y$-position were compared to FEA predictions of the gravitational couplings between the masses with the experimental conditions of Cooldown 04. 

A least-squares fitting, with $\alpha$ as a free scaling parameter, of the predicted force as a function of $y$-position to the measurements provided a best-fit $\alpha$ for a given $\lambda$. 
The best-fit $\alpha$ is based on the model that the measured force $F_m(y)$ is related to the Newtonian force $F_N(y)$ and the Yukawa force $F_Y(y)$ by
\begin{equation}
F_m(y)=F_N(y)+\alpha F_Y(y,\lambda) + F_0,
\label{fiteqn}\end{equation}
\noindent where $y$ is the $y$-position between masses and $F_0$ is some constant background. This equation was applied to the real and imaginary components of the force. MATLAB's ``fminsearch'' minimization routine was used for the least-squares fitting, with initial conditions chosen from a preliminary coarse-grained search for the fit parameters that would minimize the square error for a typical FEA result. 

\begin{figure}[t]\begin{center}
\includegraphics[width=0.7\columnwidth]{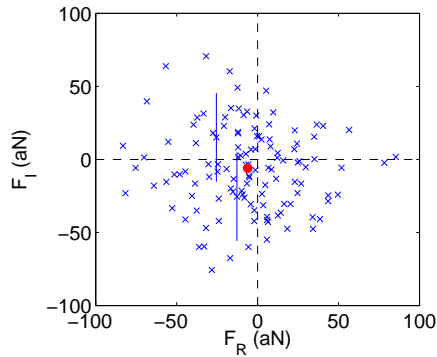}
\caption{[color online] Shown are the real and imaginary components of the Fourier transform at the frequency of interest for the second point of the Day~2 data---this is at the $y$-position at which the maximum force was measured for the day. Data here are reported in terms of the rms force, $F_I$ and $F_R$. The small blue crosses each show the result from one 30~s segment of time series data. Vertical error bars on two of these points indicate the statistical error, which is taken to be the same on each point and in all directions in this complex plane. The circle shows the mean along the two axes; the error bars on the mean are smaller than the size of the circle.\label{rawdata}}
\end{center}\end{figure}

This least squares analysis was performed separately on each day's worth of data. The real and imaginary components (rather than the magnitude and phase) of the force were considered in the fitting; the algorithm minimized the summed square difference between the measured force (considering both $\overline{F_R(y)}$ and $\overline{F_I(y)}$) and the predicted force. Though each data run had 120 measurements each of $F_R$ and $F_I$ (except the first run, which had 96), the fit was performed using the means $\overline{F_R}$ and $\overline{F_I}$ at each $y$-point, in order to reduce computing time. Tests with real and simulated data showed that the results were the same when the fit considered the entire set of data as when only the means were used. The scatter and the means of $F_R$ and $F_I$ at one $y$-point are shown in Fig.~\ref{rawdata}.

In addition to $\alpha$, four other parameters were included in the fit. An offset $y_0$ in the $y$-position accounted for the large uncertainty in the location of the range of measurements. Though the CPS gave an accurate indication of the $y$-distance between two measurement points, there was an uncertainty of $\sim 100$~$\mu$m (Tbl.~\ref{alignmenterror}) in where along the meander pattern the drive mass was located underneath the test mass. An offset in the phase $\theta_0$ was included to account for the unknown relationship between the drive signal (from the function generator) and the phase of the signal measured on the interferometer. Though FEA predictions show how the phase of a measured coupling would vary, phase offsets due to circuitry were not measured and thus the parameter $\theta_0$ was necessary in the fit.

Finally, an offset $R_0$ on the real part of the force and an offset $I_0$ on the imaginary part of the force were included in the fit. These two parameters account for the possible presence of a constant force measured in addition to any varying force (which could account for the lack of a clear periodicity in the phase, even if a measurable gravitational force was present). Tests with simulated data showed that the inclusion of these additional offset parameters did not change the other best-fit parameters by unacceptable amounts. Moreover, on the average, the best fit $R_0$ and $I_0$ would be statistically indistinguishable from 0 if no offset was included in the simulated data. The Akaike Information Criterion \cite{statsbook} indicated that using the additional offset parameters in the fit did in fact make a better model. Moreover, visual comparison of the best fit FEA curves to the data indicated that inclusion of the offset improved the fit.

\subsection{Experimental Uncertainty}
There were many experimental uncertainties to be counted in determining the best-fit $\alpha(\lambda)$. Many parameters, such as the geometry and density of the test masses, the amount of bimorph motion, and the vertical separation and tilt between the masses, entered into the determination of $\alpha$ only through the FEA and may have highly nonlinear effects on the best-fit $\alpha$. Ideally, all uncertainties would be considered in a Monte-Carlo simulation. To make the computing tractable, only $z$-separation and tilts $\theta_{xz}$ and $\theta_{yz}$ were varied in a Monte Carlo fashion. Other uncertainties were included after the least-squares results, as discussed below. Because our primary goal was to set an upper bound on deviations from Newtonian gravity, when simplifications were required, we chose to use ``worst-case'' estimates if possible, including errors in a way that would make the best-fit $\alpha(\lambda)$ larger. 

\subsection{Monte Carlo}
The FEA simulation was run 320 times, with the input parameters of the $z$-separation, $\theta_{xz}$, and $\theta_{yz}$ varied about their respective experimental best-guess values. For each FEA run, the values of these three parameters were chosen at random from Gaussian distributions with means equal to the respective best-guess values and standard deviations equal to the experimental uncertainties, as shown in Tbl.~\ref{alignmenterror}. The range of the $z$-separation was considered part of the uncertainty in $z$; the $z$-value was varied about the middle of this range (26~$\mu$m) and half of the range (2~$\mu$m) was added in quadrature to the experimental uncertainty. Eight values of $\lambda$ were chosen for the FEA models. 

\subsection{Unvaried Inputs to the FEA}
In the FEA model, the mass densities of gold and silicon were taken to be the bulk densities. Though the test mass was rotated in the plane of the cantilever, this rotation was not considered in the FEA model; rotation of the test mass would only affect the results at the percent level (and would lower $\alpha(\lambda)$). The test mass was modeled as a prism ($50\times 50\times 30$)~$\mu$m$^3$, with a symmetric pattern of  52 blocks on the bottom face to approximate the curved section of the test mass. This model underestimates the size of both the prism part and the rounded part of the test mass (as compared to what is given in Tbl.~\ref{masserror}); since this underestimate will only increase the final $\alpha(\lambda)$ bound, the approximation is acceptable. The test mass was taken to be tilted at 0.35~rad from the horizontal plane of the cantilever. The model of the test mass in the FEA code did not account for the recessed face discussed in Sec.~\ref{tmsec}. Considering the exponential dependence of the Yukawa potential, this missing volume (on the side face tilted away from the cantilever) would only affect the calculated gravitational force at the $2\%$ level. This level of error is small compared to other uncertainties in $\alpha$ and may be ignored in this analysis.

The $x$-position between the drive mass and the test mass was taken to be the experimental best-guess value. Both the Newtonian and the Yukawa gravitational forces do depend slightly on the $x$-position. However, this dependence changes with the tilt $\theta_{xz}$ and there was no clear ``worst-case'' value for $x$-position to use in the FEA model. To reduce computing time, the model computed only the force for a range of 200~$\mu$m in $y$-position. The dc force for this reduced range was mirrored to make a range of 600~$\mu$m from which the ac force was derived, a simplification that affected the best-fit $\alpha$ at the $10\%$ level. This range of $y$-positions was centered near the best-guess center of the range of data-acquisition and the same FEA model was used to fit to each day of data. Due to possible tilt of the drive mass, fitting each day of data to the same $y$-range could also incur errors on the order of $10\%$ on the best-fit $\alpha(\lambda)$.

\subsection{Uncertainty in Bimorph Amplitude}
A bimorph amplitude of 125~$\mu$m was considered in the FEA model. To account for the experimental uncertainty of 10~$\mu$m in this factor, each Monte Carlo result was also fit to the data for bimorph amplitudes in the range of 95--155~$\mu$m (in steps of 5~$\mu$m), accounting for three times the experimental uncertainty. With the dependence of $\alpha$ on the bimorph amplitude thus determined, each best-fit value of $\alpha$ for a bimorph amplitude of 125~$\mu$m was varied 50 times according to a random sampling from a Gaussian distribution of bimorph amplitudes, yielding a set of $320\times 50$ best-fit values of $\alpha$ for each chosen $\lambda$ value. Different samplings of bimorph amplitudes (from the Gaussian) were used for each day of data.

\subsection{Uncertainties in Multiplicative Factors}
Several factors in the FEA model were not varied as Monte-Carlo parameters. Most of these parameters entered as multiplicative factors in the determination of the best-fit $\alpha(\lambda)$. 

Though the model on which the fit is based (Eq.~\ref{fiteqn}) considers measurements at each $y$-point, due to thermal noise, data will never match FEA predictions at the predicted zero minima of the force magnitude. Thus, the best-fit $\alpha$ is mostly determined by the quality of the fit of the data to the maxima of the prediction curves. Roughly, $\alpha=(F_m-F_N)/F_Y$, where only the maxima of the measured force ($F_m$), the Newtonian force ($F_N$), and the Yukawa force ($F_Y$) are considered. This argument holds true in the presence of a small offset force. In this experiment, $F_m\gg F_N$ and thus, the proportional error in $\alpha$ was determined by a quadrature sum of the proportional errors in $F_Y$ and $F_m$: $(\delta \alpha/\alpha)^2 \sim (\delta F_m/F_m)^2+(\delta F_Y/F_Y)^2$, where $\delta \alpha$ is the uncertainty in the best-fit $\alpha$ resulting from the uncertainties $\delta F_m$ and $\delta F_Y$. This is true for small ($<10\%$) relative uncertainties in $F_Y$ and even for large ($\sim$ 50$\%$) relative uncertainties on $F_m$. The uncertainty in $F_N$ was ignored because the contribution of $F_N$ in this equation is small. 

The uncertainties in the measured force (listed in Tbl.~\ref{forceerror}) were all multiplicative factors, which would scale the entire curve of $F_m(y)$ up or down. Uncertainties in the respective densities of the masses were multiplicative factors in $F_Y$. However, all densities were taken to be the given bulk densities. Voids and uncertainties in the shape of the test mass all entered, to first order, as multiplicative factors in $F_Y$. The indeterminate composition areas at the bottom of the drive mass were also approximated as a very small multiplicative factor on $F_Y$; FEA models showed that even decreasing the height of the drive mass from 100~$\mu$m to 90~$\mu$m had at most a few percent effect on the magnitude of $F_Y$ for the studied values of $\lambda$. Together, these factors summed to $8.4\%$ relative uncertainty on $F_Y$. 

To account for these uncertainties in the multiplicative factors of $F_m$ and $F_Y$, each of the prediction curves and the data curves could be varied by Gaussian distributions representing the uncertainty of the respective multiplicative factors and the best-fits could be sought between the new, much larger sets of prediction and data curves. This parametric bootstrap was simplified because, as argued above, the best-fit $\alpha$ changed with scalings of $F_m(y)$ and $F_Y(y)$ in a predictable way. Thus, the uncertainties in these multiplicative factors were counted by varying the best-fit $\alpha(\lambda)$ values according to the relative uncertainties in $F_m$ and $F_Y$. 

\subsection{Statistical Uncertainty}
Though a least-squares fit does find a best fit that accounts for the possibility of fluctuations in the data about the ``true'' values, the statistical uncertainty in the measurements does lead to an error in the best-fit results. In the simple case of a least-squares linear fit, for example, this uncertainty in the fit parameters can be easily calculated \cite{barlow}. In this case, the uncertainty in the best-fit results cannot be determined analytically. To determine the uncertainty in $\alpha$ resulting from the statistical uncertainty (due to thermal noise) on each measurement, the measurements were artificially varied by the statistical uncertainty found in the data. At a given $y$-point, an offset, drawn at random from a Gaussian distribution with a mean of 0 and a standard deviation equal to the standard error on $\overline{F_R}$, was added to the measured $\overline{F_R}$. The measured $\overline{F_I}$ was similarly perturbed. At each $y$-point, the measurements were dithered in this way, adding to the measured points their statistical uncertainty. The results of a typical FEA simulation were then fit to these dithered points. The process was repeated multiple times and from this, the uncertainty in $\alpha$ due to the statistical uncertainty in the data was found to be $22\%$. This compared well to the standard deviation of the best-fit $\alpha$ over many sets of simulated data.

\subsection{Summation of Errors}
Uncertainties in the most important geometric factors (vertical separation and tilt angles) were considered in the Monte Carlo run. Uncertainties in the bimorph amplitude were considered by varying the best-fit $\alpha(\lambda)$ results. Uncertainties in the multiplicative factors and the statistical uncertainty in the best-fit results were considered by varying the best-fit results once more. In this case, each best-fit result was multiplied by a set of 50 random samples from a Gaussian of mean 1 and a standard deviation equal to the quadrature sum of the uncertainties due to the multiplicative factors and the statistical uncertainty on $\alpha$. These factors are summarized in Tbl.~\ref{alphaerrors}. Finally, for each value of $\lambda$, there were 800 000 best-fit $\alpha$ results. 

        \begin{table}[t]
        \begin{center}
        \caption{Errors on best-fit $\alpha$}
        \smallskip
        \begin{tabular} { l  r  r  r}
        \hline
        \hline
        Parameter & $\%$ Uncertainty\\
        \hline
	Volume of test mass	&7.6\\
	Voids in test mass	&2.5\\
	Voids in drive mass	&2.5\\
	Drive mass polished Au/Si boundary	&1.0\\
	\hline
	 \textbf{\textit{Total}} Multiplicative error in $F_Y$	&8.4\\
	\hline
	Multiplicative error on measured force	&17\\
	Statistical error	&22\\
	\hline
	 \textbf{\textit{Total}} {\bf Uncertainty in $\alpha$}	&29\\
	\hline
        \hline
        \end{tabular}
        \label{alphaerrors}
        \end{center}
        \end{table}

\subsection{Best-Fit Results}
\begin{figure}
\includegraphics[width=\columnwidth]{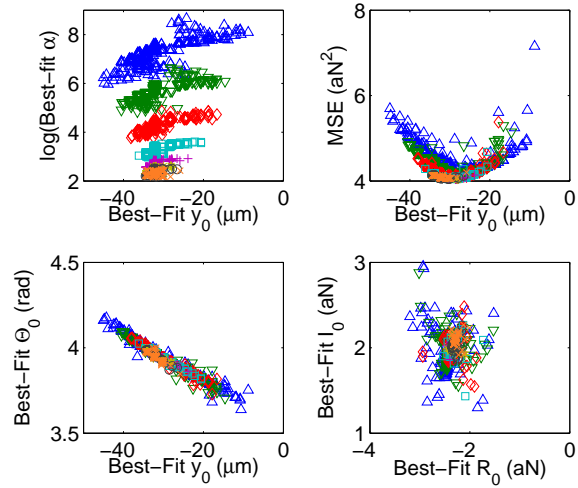}
\caption{[color online] Correspondence between parameters for the fits of the Monte Carlo results to Day~1 data. In the upper left, the best-fit $\alpha$ values are shown; from the top (larger $\alpha$), the groups of best-fit values correspond to values of $\lambda$ of 4, 6, 10, 18, 34, 66, 130, and 258~$\mu$m. Different symbols are used for each value of $\lambda$, with the same symbols used in each plot. Plots show the mean square error (MSE) for the fit and the best-fit parameters $y_0$, $\theta_0$, $R_0$, and $I_0$. \label{fitparams}}
\end{figure}

Best-fit $y_0$, $\theta_0$, $R_0$, and $I_0$ in these fits have no important physical meaning. However, examination of the best-fit results for these other fit parameters was useful in evaluation of the analysis method and demonstration of the robustness of the fitting procedure. Fig.~\ref{fitparams} shows correlations among these parameters and the mean square error from the best-fit. As expected, the sets of best-fit $\alpha$ for each value of $\lambda$ showed an exponential dependence on the effective vertical separation (including both $z$ and the additional separation resulting from $\theta_{xy}$ and $\theta_{yz}$) between the masses.  

The mean best-fit $y_0$ for each of the two data days indicated $\sim 40$~$\mu$m difference (for a given CPS reading of $y$-position) between the two days. The best fit $\theta_0$ varied by almost $\pi$ between the two data days; this is the expected change if there were indeed a gravitational signal, given the difference in $y_0$. The magnetic experiment of Cooldown~02 showed that a shift of this size in the CPS reading from one day to the next could be expected. Moreover, if the measured force was a non-gravitational background (such as electrical noise), an arbitrary shift in the best-fit $y_0$ between data days could be expected. 

\begin{figure}
\includegraphics[width=\columnwidth]{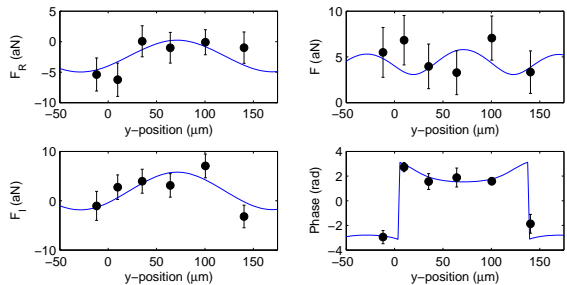}
\caption{[color online] The measured force as $\overline{F_R}$, $\overline{F_I}$, magnitude, and phase for Day~1 as compared to a typical best-fit Monte Carlo result, for $\lambda=18$~$\mu$m. The force magnitude $F$ is $\sqrt{\overline{F_R}^2+\overline{F_I}^2}$. Error bars show the statistical error on the mean.\label{bestfitex}}
\end{figure}

\begin{figure}
\includegraphics[width=\columnwidth]{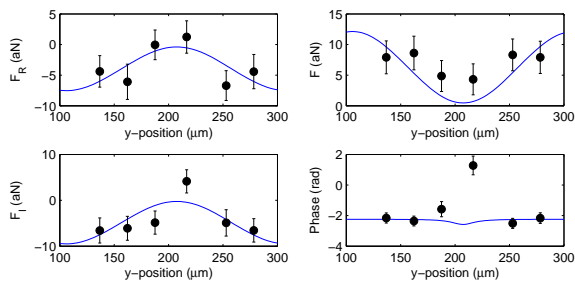}
\caption{[color online] The same information as in Fig.~\ref{bestfitex}, except for Day~2 data.\label{bestfitex2}}
\end{figure}

Scaled and shifted by the best-fit parameters, the FEA results were almost all within two standard deviations of the measured $\overline{F_R}$, $\overline{F_I}$, force magnitude $F$, and measured phase with respect to the drive. A typical best-fit result, in comparison to the measurements, is shown in Figs.~\ref{bestfitex} and \ref{bestfitex2}.

For each of the two data days, the mean of the best-fit offset force (described by $R_0$ and $I_0$) had a magnitude close to the mean magnitude of the measured force across the day's $y$-scan. The magnitude of the mean best-fit offset for Day~1 was $3.1\times 10^{-18}$~N, at a phase of 2.4~rad. For Day~2,  the mean best-fit offset had a magnitude of $6.3\times 10^{-18}$ N, at a phase of -2.3~rad.

The measured force on Day~2 was larger than the measured force on Day~1; correspondingly, the typical best-fit $\alpha$ for Day~2 was $20\%$--$40\%$ higher than the best-fit $\alpha$ for Day~1. The mean square errors for the fits on Day~2 were $20\%$ (at small $\lambda$) to $40\%$ (at larger $\lambda$) higher than the errors on the same fits for Day~1. 
 
Though the parameter $y_0$ accounted for some of the uncertainty in the location of the range over which measurements were made, the least squares fit did not account for uncertainty in the separation between $y$-points. Uncertainty in the $x$-position between masses was also not considered. Furthermore, the possibility of scaling errors in $F_m$ that changed over the course of the day (such as a continual drift of $f_0$ away from $3f_d$ over the course of the day, changing both the scaling and the phase of any measured force) were not included in the analysis. Inclusion of these effects, all expected to be relatively small, has been left for future work.

\subsection{Interpretation of the Results}
The Monte Carlo results and subsequent varying of the best-fit results provide a spread of $\alpha$ for each $\lambda$ considered. These results may point to a force with a finite magnitude that shows gravity-like features in our experiment. However, due to the small size of the measured force as compared to thermal noise, we are not able to determine at this point whether the measured force originated from a gravitational coupling between the drive and test masses. If the above results were to be interpreted as a signature for a true force, the uncertainties would be considered in the opposite sense (with respect to the best-guess values) than for this analysis; this would yield smaller values of $\alpha(\lambda)$, a lower bound. Such an approach will not be explored in this paper since more data with a better signal-to-noise ratio is needed. Future work to this end is discussed in Sec.~\ref{futurework}.  Furthermore, to confirm the existence of a gravitational force, measurements would have to be made as a function of the mass or the separation between masses in the experiment. 

In the meantime, we will use our current results to put new upper bounds on Yukawa-type deviations from Newtonian gravity. This is described in the next section.  

\section{Drawing a Bound on $\alpha(\lambda)$}
\subsection{Summary of Procedures} 
The $\alpha(\lambda)$ bound was derived from a measurement of cantilever motion, as described in the previous sections. The drive mass was oscillated underneath the cantilever bearing the test mass. An interferometric measurement of cantilever motion was recorded as time-series data. The time-series data from the interferometer were Fourier-transformed, averaged, and compared to the drive signal in order to determine how much the cantilever was moving at the frequency of interest (Eq.~\ref{vppeqn}). From this motion, the force on the cantilever was deduced, as described in Eq.~\ref{acforceeqn}. The equilibrium position of the drive mass with respect to the test mass was varied longitudinally and measurements were recorded as a function of this position; room temperature alignment and capacitive sensors used during the low-temperature measurement indicated the relative position between masses. The measured force as a function of the drive mass position was fit to a model of the gravitational force (including a Yukawa force) between masses; this model was based on FEA using Eqs.~\ref{FEAnewton} and \ref{FEAyuk}. Experimental uncertainties were considered via Monte Carlo variation of the inputs to the FEA and variation of the results of the fit. The best fit between the measurements and each FEA result yielded a set of best-fit $\alpha(\lambda)$ for a Yukawa potential that could be consistent with our data. The final upper-bound results were determined from this set of best-fit $\alpha(\lambda)$.

\subsection{Determination of the Final Results}
\begin{figure}
\includegraphics[width=0.7\columnwidth]{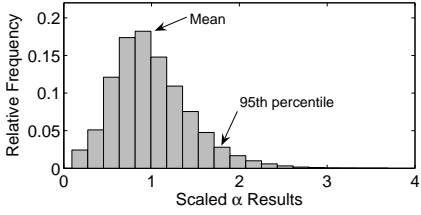}
\caption{[color online] An example histogram of the fully-varied best-fit $\alpha$ results for $\lambda=18$~$\mu$m and Day~1 data. The Scaled $\alpha$ is $\alpha/\overline{\alpha}$ and the Relative Frequency is scaled by the total number (800,000) of varied best-fit $\alpha$ values. \label{alphahisto}}
\end{figure}

The goal is to set an upper-bound on Yukawa-type deviations from Newtonian gravity at the $95\%$ confidence level. Though all experimental uncertainties were assumed to be Gaussian, the resulting distribution of $\alpha$ for each $\lambda$ was not; especially in the case of small $\lambda$, where the best-fit $\alpha$ had a highly nonlinear dependence on the vertical separation and tilts between masses, the distributions were very asymmetric. No assumption was made of the analytic form of the distribution; the 95th percentile $\alpha$ from the distribution provided the result at the desired (one-sided) confidence level, as diagrammed in Fig.~\ref{alphahisto}. The 95th percentile $\alpha(\lambda)$ after the variations of the Monte Carlo results was found to be less than $35\%$ more than the 95th percentile $\alpha(\lambda)$ from the Monte Carlo variation alone. 

         \begin{table}[t]\begin{center}
        \begin{minipage}{\columnwidth}\begin{center}
        \caption[Summary of results]{Summary of results}
        \smallskip
        \begin{tabular} { l  r  r  r}
        \hline
         {\it Day~1}	&	&\\
         \hline
         $\lambda$ ($\mu$m)	&MC Mean $\alpha$\footnote{This is the mean of the best-fit $\alpha$ results from fitting the set of Monte Carlo-varied FEA prediction curves to data.} &Full Set Mean $\alpha$\footnote{This includes the variation of Monte Carlo results to account for uncertainty in bimorph motion and the statistical uncertainty in $\alpha$ combined with experimental uncertainty in the multiplicative factors.}	&Upper Bound $\alpha$\footnote{This is the $95\%$ confidence level result.}\\
         \hline
	4		&$3.6\times 10^7$	&$3.8\times 10^7$	&$1.5\times 10^8$ \\
	6		&$6.7 \times 10^5$	&$6.9\times 10^5$	&$2.0 \times 10^6$\\
	10		&$2.2 \times 10^4$	&$2.3\times 10^4$	&$5.1 \times 10^4$\\
	18		&$2.0 \times 10^3$	&$2.1\times 10^3$	&$3.7 \times 10^3$\\
	34		&$5.0 \times 10^2$	&$5.1\times 10^2$	&$8.4 \times 10^2$\\
	66		&$2.7 \times 10^2 $	&$2.8\times 10^2$	&$4.4 \times 10^2$\\
	130		&$2.2 \times 10^2$	&$2.3\times 10^2$	&$3.6 \times 10^2$ \\
	258		&$2.1 \times 10^2$	&$2.1\times 10^2$	&$3.3 \times 10^2$\\
        \hline
        \hline
         {\it Day~2}	&	&\\
         \hline
         $\lambda$ ($\mu$m)	&MC Mean $\alpha$ &Full Set Mean $\alpha$	&Upper Bound $\alpha$\\
	\hline
	4		&$5.1\times 10^7$	&$5.2\times 10^7$	&$2.0\times 10^8$ \\
	6		&$8.9 \times 10^5$	&$9.1\times 10^5$	&$2.8 \times 10^6$\\
	10		&$2.9 \times 10^4$	&$3.0\times 10^4$	&$6.8 \times 10^4$\\
	18		&$2.5 \times 10^3$	&$2.6\times 10^3$	&$4.8 \times 10^3$\\
	34		&$6.2 \times 10^2$	&$6.4\times 10^2$	&$1.1 \times 10^3$\\
	66		&$3.4 \times 10^2 $	&$3.5\times 10^2$	&$5.7 \times 10^2$\\
	130		&$2.7 \times 10^2$	&$2.8\times 10^2$	&$4.6 \times 10^2$ \\
	258		&$2.5 \times 10^2$	&$2.6\times 10^2$	&$4.2 \times 10^2$\\
	\hline
	\hline
        \end{tabular}
        \label{resultstable}
        \end{center}
        \end{minipage}
                \end{center}
        \end{table}

Though this analysis assumed $\alpha>0$, because the measured force was $\sim 200$ times greater than the Newtonian gravitational force in this case, the results for $\alpha<0$ would be statistically indistinguishable (when comparing $\left |\alpha \right |$) from the results for $\alpha>0$.  Thus, our results may be considered as a bound on $|\alpha|$.

\begin{figure}
\includegraphics[width=\columnwidth]{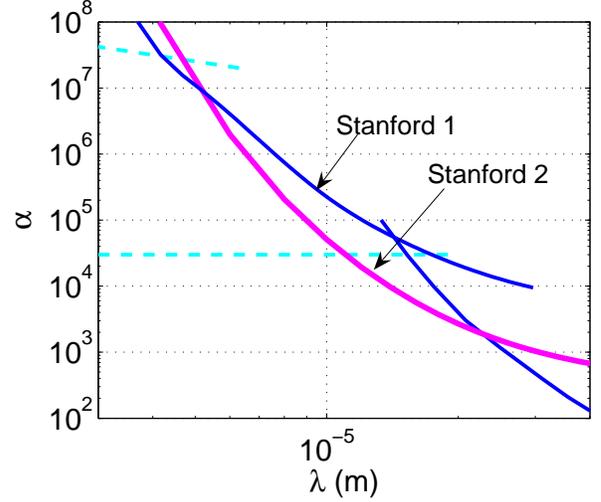}
\caption{[color online] The parameter space of $\alpha$-$\lambda$ of Yukawa-type deviations from Newtonian gravity. Results from this paper (Stanford 2) are shown in comparison to previous results from this experiment \cite{frogs} (Stanford 1). Also included are theoretical predictions (dashed line) and an experimental result (solid line) as shown Fig.~\ref{paramspace}. \label{phspresults}}
\end{figure}

Results for Day~1 were different from results for Day~2. As discussed above, there was indication of a larger non-gravitational background on Day~2 as compared to Day~1. This could be the cause of the slightly higher best-fit $\alpha(\lambda)$ on Day~2. If the difference between the mean best-fit $\alpha(\lambda)$ and the 95th percentile $\alpha(\lambda)$ is considered to be two standard deviations, then the difference between Day~1 and Day~2 results is at an acceptable level. The Day~1 results are plotted in Fig.~\ref{phspresults} and taken as our $95\%$ confidence-level bound. Results from Day~1 were chosen because the error on the fit and diagnostics from the data (such as signal levels at $f_d$) suggested that Day~1 measurements suffered from a smaller non-gravitational background.

These results are properly interpreted as a bound on, not a discovery of, deviations from Newtonian gravity. The choice of the Yukawa parameterization is appropriate though not all-inclusive; results could be similarly analyzed using other parameterizations (such as a power-law potential) for deviations from Newtonian gravity. We are $95\%$ confident that no Yukawa-type gravitational potential exists with $\alpha(\lambda)$ above the bound reported for the Day~1 results in Tbl.~\ref{resultstable}. 

\section{Conclusions}
In this paper, we have described our experimental apparatus and the latest data it has produced. While this experiment was based on the same principles as the one that yielded our first results \cite{frogs}, it incorporated many substantial improvements in the design, technique, and data analysis. The results presented here represent almost an order of magnitude improvement over previous results at $\lambda\sim 20$~$\mu$m \cite{frogs}, yielding the most stringent experimental constraints on Yukawa-type deviations from Newtonian gravity at length scales of 6--20~$\mu$m. This new bound provides constraints on predictions of moduli and gauge bosons. These new constraints do not rule out string theory or supersymmetry or the possibility of large extra dimensions. However, the $\alpha(\lambda)$ bound limits what kind of particles could be included in any of these theories of physics beyond the standard model.

\section{Future Prospects}\label{futurework}
These results are a quantitative and qualitative improvement over previous results. Moreover, our apparatus and analysis are more complete and robust, providing a strong foundation on which to build future experiments to test gravity even further. Reducing electrical and mechanical coupling between the bimorph and the interferometer and using a test mass without a curved face (to improve fiber-mass alignment, increasing fringe height) could increase the signal-to-noise by an order of magnitude. It is unlikely that the thermal noise limit in this experiment will be reduced by more than a factor of two. However, measuring with a test mass not tilted with respect to the cantilever would increase the expected gravitational force and could easily yield a factor of two improvement on the $\alpha(\lambda)$ bound. A new cantilever design may provide a magnetic calibration, which will greatly reduce uncertainty in $y$-position and allow for a more precise fitting of results to FEA predictions. We are developing a similar apparatus with a circular drive geometry, which we expect to be at least an order of magnitude more sensitive than the current one.

\acknowledgments
We are grateful to the NSF for funding (grant PHY-0244932). We thank Dan Rugar for many helpful discussions about experimental techniques and Savas Dimopoulos for providing the initial theoretical motivation for our work.


\begin{thebibliography}{99}

\bibitem{add1}
N.~Arkani-Hamed, S.~Dimopoulos, and G.~Dvali, Phys.~Lett.~B {\bf 429}, 263 (1998).

\bibitem{add2}
N.~Arkani-Hamed, S.~Dimopoulos, and G.~Dvali, Phys.~Rev.~D {\bf 59}, 086004 (1999).

\bibitem{sundrum}
R.~Sundrum, J.~High Energy Phys. {\bf 07}, 001 (1999).

\bibitem{savasmoduli}
S.~Dimopoulos and G.~F.~Giudice, Phys.~Lett.~B {\bf 379}, 105 (1996).

\bibitem{beane}
S.~Beane, Gen.~Relativ.~Gravit. {\bf 29}, 945 (1997).

\bibitem{savasradion}
I.~Antoniadis, S.~Dimopoulos, and G.~R.~Dvali, Nucl.~Phys. {\bf B516}, 70 (1998).

\bibitem{ignatios}
I.~Antoniadis, N.~Arkani-Hamed, S.~Dimopoulos, and G.~R.~Dvali, Phys.~Lett.~B {\bf 436}, 257 (1998).

\bibitem{dilaton}
D.~B.~Kaplan and M.~B.~Wise, J.~High Energy Phys. {\bf 08}, 037 (2000).

\bibitem{frogs}
J.~Chiaverini, S.~J.~Smullin, A.~A.~Geraci, D.~M.~Weld, and A.~Kapitulnik, Phys.~Rev.~Lett. {\bf 90}, 151101 (2003).

\bibitem{jacthesis}
J.~Chiaverini, Ph.D. thesis, Stanford University (2002).

\bibitem{SJSthesis}
S.~J.~Smullin, Ph.D. thesis, Stanford University (2005).

\bibitem{SJSmoriond}
S.~J.~Smullin, D.~M.~Weld, A.~A.~Geraci, J.~Chiaverini, and A.~Kapitulnik, in {\it Gravitational Waves and Experimental Gravity, Proceedings of the XXXVIIIth Rencontres de Moriond}, edited by J.~Dumarchez and J.~T.~T.~V\^an (The Gioi Publishers, Vietnam, 2003), pp.~243-250.

\bibitem{SJSslac}
S.~J.~Smullin, A.~A.~Geraci, D.~M.~Weld, and A.~Kapitulnik, in {\it Proceedings of the SLAC Summer Institute on Particle Physics (SSI04), Menlo Park, 2004}, edited by J.~Hewett, J.~Jaros, T.~Kamae, and C.~Prescott, eConf C040802 (2004).

\bibitem{lamoreaux}
S.~K.~Lamoreaux, Phys.~Rev.~Lett. {\bf 78}, 5 (1997).

\bibitem{pricenature}
J.~C.~Long, H.~W.~Chan, A.~B.~Churnside, E.~A.~Gulbis, M.~C.~M.~Varney, and J.~C.~Price, Nature {\bf 421}, 922 (2003).

\bibitem{adelberger04}
C.~D.~Hoyle, D.~J.~Kapner, B.~R.~Heckel, E.~G.~Adelberger, J.~H.~Gundlach, U.~Schmidt, and H.~E.~Swanson, Phys.~Rev.~D {\bf 70}, 042004 (2004).

\bibitem{sarid}
D.~Sarid, {\it Scanning Force Microscopy with Applications to
Electric, Magnetic and Atomic Forces} (Oxford University
Press, 1994), revised ed.

\bibitem{E_Si_size}
X.~Li, T.~Ono, Y.~Wang, and M.~Esashi, App.~Phys.~Lett., {\bf 83}, 3081(2003).

\bibitem{snf_nitridestress}
M.~H.~Badi and E. Wong, Stanford Nanofabrication Facility, LPCVD Stress Measurements--Nitride,
retrieved June 11, 2005. \url{http://snf.stanford.edu/Equipment/tylanlpcvd/FilmStress.html}.

\bibitem{sin_stress_a}
A.~Kaushik, H.~Kahn, and A.~H.~Heuer, J.~Microelectromech.~Syst. {\bf 14}, 359 (2005).

\bibitem{sin_stress_b}
Y.~Toivola, J.~Thurn, and R.~F.~Cook, J.~Appl.~Phys. {\bf 94}, 6915 (2003).

\bibitem{cirlex}
Fralock Industries, \url{http://www.cirlex.com}, 21054 Osborne Street, Canoga Park, CA 91304, USA.

\bibitem{cps}
S.~B.~Field and J.~Barentine, Rev.~Sci.~Instrum. {\bf 71}, 2603 (2000).

\bibitem{rugarintf}
D.~Rugar, H.~J.~Mamin, and P.~Guethner, Appl.~Phys.~Lett. {\bf 55}, 2588 (1989).

\bibitem{rugar_ultramic}
T.~R.~Albrecht, P.~Gr\"{u}tter, D.~Rugar, and D.~P.~E.~Smith, Ultramicroscopy {\bf 42--44}, 1638 (1992).

\bibitem{stiction}
A broad discussion of this stiction effect is given in R.~Maboudian and R.~T.~Howe, J.~Vac.~Sci.~Technol.~B {\bf 15}, 1 (1997).

\bibitem{laura}
P.~A.~A.~Laura, J.~L.~Pombo, and E.~A.~Susemihl, J.~Sound Vib. {\bf 37}, 161 (1974).

\bibitem{rugar_pc}
D.~Rugar, Private communication.

\bibitem{statsbook}
K.~P.~Burnham and D.~R.~Anderson, {\it Model Selection
and Inference: A Practical Information-Theoretic Approach} (Springer, New York, 1998).

\bibitem{barlow}
R.~J.~Barlow, {\it Statistics: A Guide to the Use of Statistical Methods in the Physical Sciences} (John Wiley and Sons,1989).

\bibitem{lamoreauxalphlam}
M.~Bordag, B.~Geyer, G.~L.~Klimchitskaya, and V.~M.~Mostepanenko, Phys.~Rev.~D {\bf 58}, 075003 (1998).

\bibitem{andysavas}
S.~Dimopoulos and A.~A.~Geraci, Phys.~Rev.~D {\bf 68}, 124021 (2003).


\end{thebibliography}
\end{document}